\documentclass[aps,pra,showpacs,floatfix,reprint]{revtex4-1}

\usepackage{amsfonts}
\usepackage{amsmath}
\usepackage{amssymb}
\usepackage{latexsym}
\usepackage{graphicx}

\begin{document}

\title{Generalized multi-terminal decoherent transport: Recursive algorithms
and applications to SASER and giant magnetoresistance.}
\author{Carlos J. Cattena$^{1}$ Lucas J. Fern\'andez-Alc\'azar$^{1}$, Ra\'ul
A. Bustos-Mar\'un$^{1,2}$, Daijiro Nozaki$^{3}$ and Horacio M. Pastawski$%
^{1} $}

\begin{abstract}
Decoherent transport in mesoscopic and nanoscopic systems can be formulated
in terms of the D'Amato-Pastawski (DP) model. This generalizes the Landauer-B%
\"{u}ttiker picture by considering a distribution of local decoherent
processes. However, its generalization for multi-terminal setups is lacking.
We first review the original two-terminal DP model for decoherent transport.
Then, we extend it to a matrix formulation capable of dealing with
multi-terminal problems. We also introduce recursive algorithms to evaluate
the Green's functions for general banded Hamiltonians as well as local
density of states, effective conductances and voltage profiles. We finally
illustrate the method by analyzing two problems of current relevance. 1)
Assessing the role of decoherence in a model for phonon lasers (SASER). 2)
Obtaining the classical limit of Giant Magnetoresistance from a
spin-dependent Hamiltonian. The presented methods should pave the way for
computationally demanding calculations of transport through nanodevices,
bridging the gap between fully coherent quantum schemes and semiclassical
ones.
\end{abstract}

\maketitle

\affiliation{$^{1}$Instituto de F\'{\i}sica Enrique Gaviola and Facultad de Matem\'{a}tica 
Astronom\'{\i}a y F\'{\i}sica, Universidad Nacional de C\'{o}rdoba,
Ciudad Universitaria, C\'{o}rdoba, 5000, Argentina,} 
\affiliation{$^{2}$Facultad de Ciencias Qu\'{\i}micas, Universidad Nacional de C\'{o}rdoba,
Ciudad Universitaria, C\'{o}rdoba, 5000, Argentina,} 
\affiliation{$^{3}$Institute for Materials Science and Max Bergmann Center of
Biomaterials, Dresden University of Technology, D-01062 Dresden, Germany}

\section{Introduction}

Quantum transport at the nanoscale \cite{DiCarlo2004,Zimbovskaya2011,Cunibertibook} is a
blooming field where the properties of matter can be explored in a realm
where quantum effects become crucial. In particular, the control of quantum
interference phenomena and their interplay with the electronic structure
offers a fascinating opportunity to overcome some of the usual constraints
of our macroscopic classical world. \cite%
{Nature2001,Nature2012,Nitzan2003,Bustos2013,Rickhaus} However, at the
nanoscale, both quantum \textit{and} classical behavior can be expected.
This last emerges from the unavoidable environmental degrees of freedom. 
\cite{Ratner2013} An exciting example of the competition among those
behaviors is electron-transfer in natural and artificial photosynthesis.
There, the interplay between localizing interferences and environmentally
induced decoherence seems to have a fundamental role in optimizing excitonic
transfer. \cite{Huelga2008,Lloyd2009} This phenomenon falls in line with
what is known in low dimensional conductors. Indeed, transport properties of
highly ordered 1-D systems is determined by the fast quantum diffusion of
local excitations, and thus become weakened by decoherence. On the other
hand, in disordered 1-D wires, quantum coherence allows the destructive
interferences that produce electronic localization. While these phenomena
are roughly described by introducing imaginary energies in the Kubo
formulation, it is at the cost of overlooking charge conservation. \cite%
{Thouless-Kirkpatrick}

Landauer's picture has almost no rival in what concerns to electronic
coherent transport.\cite{Landauer1999} In its simplest form, conductance is
determined by the transmission probability (either quantum or classical)
among electrodes. Paradoxically, quantum transmittance is much simpler to
evaluate than its classical counterpart. Thus, the great majority of work
focus on the evaluation of the coherent transmittance setting aside
incoherent processes. An extension of this approach, developed by Markus B%
\"{u}ttiker,\cite{Buttiker1986} applies the Kirchhoff laws to a system
connected to multiple terminals. This allows to consider different voltage
probes as well as multiple current sources and drains. The self-consistent
non-equilibrium chemical potentials at the voltmeters must ensure current
cancellation. The resulting transport coefficients fulfill the Onsager's
reciprocity relations. Additionally, B\"{u}ttiker had the crucial insight%
\cite{Buttiker1986PRB} that a voltage probe implies a classical measurement
and thus it acts as a decoherence source. This concept was further
formulated by D'Amato and Pastawski introducing a Hamiltonian description 
\cite{Damato-Pastawski} (henceforth the DP model). In this description, the
decoherent local probes can be assimilated to incoherent scattering by
delta-function potentials\cite{GLBE1,GLBE2}. This is founded in the Keldysh,
Kadanoff and Baym's quantum fields formalism\cite{Danielewicz1984} for the
non-equilibrium Green's functions. \cite%
{Danielewicz1984,Kadanoff1989,Rammer1986} There, the integro-differential
equations are simplified by evaluating the currents and chemical potentials
in a linearized scheme that involves a matrix containing only transmittances
among different points in the sample. The DP model also provides a compact
solution for an arbitrary distribution of incoherent local scattering
processes. These lead to a momentum relaxing decoherence that produces
diffusion and a further increase in the resistance. The final set of linear
equations relate the local chemical potentials and the currents through a
transmittances matrix. \cite{Datta90} This results in the Generalized
Landauer-B\"{u}ttiker Equations\ (GLBE) that solve the DP model.

The original presentation of the DP model is constrained to two terminal
problems. Thus, in spite of the growing need to include the effects of
decoherent processes,\cite{Maassen2009, Horvat2013} its applications
remained mostly reduced to a few one-dimensional problems. \cite%
{Zimbovskaya2002,Zwolak2002,Gagel1996,Nozaki2008,PAni-CBP,Nozaki2012,Anantram2013}
Besides, since the method deals with a great number of self-consistent local
chemical potentials, it often involves a cumbersome matrix inversion. Thus,
a general multi-terminal formulation of the DP model for decoherent
transport and an efficient computational strategy are still lacking.

In this paper we generalize the D'Amato-Pastawski model for multi-terminal
problems, presenting a decimation-based method for the calculation of the
decoherent conductance. In Sec. \ref{sec:BasicTools} we introduce the basic
tools, based on a decimation procedure that yields the parameters of an
effective Hamiltonian. In Sec. \ref{sec:DP} we overview the original DP
model. In Sec. \ref{sec:Computational}, we generalize the DP model for
multi-terminal setups. We also provide a recursive algorithm for the
calculation of Green's functions of general banded Hamiltonians. Then, we
show two application examples. In Sec. \ref{sec:example1} we consider a
simple model of a phonon-laser (SASER) based on the electron-phonon
interaction in a quantum dot \cite{Kent} where we asses the role of
decoherence in the SASER efficiency. In Sec. \ref{sec:example2}, we consider
the spin dependent electronic transport in a ferromagnetic wire where the
Giant Magnetoresistance (GMR) \cite{Fert08} shows up. We show that our
formulation describes the complete cross-over from a quantum transport to
the GMR semiclassical regime. In Sec. \ref{sec:conclusions} we summarize our
results and conclude that our formulation can handle decoherent transport in
a wide variety of problems beyond the typical two-terminal calculations.

\section{\label{sec:BasicTools}Decimation Procedures and Effective
Hamiltonians}

Even the simplest quantum devices involve a huge number of degrees of
freedom and thus their study can not be carried out without proper
simplifications. For example, a tight-binding Hamiltonian describing a
device or molecule with $N$ states (or orbitals) is, \cite{Pastawski-Medina} 
\begin{equation}
\hat{H}_{S}=\sum\limits_{i=1}^{N}\left\{ E_{i}\hat{c}_{i}^{\dagger }\hat{c}%
_{i}^{{}}+\sum\limits_{\substack{ j=1  \\ (j\neq i)}}^{N}\left[ V_{i,j}\hat{c%
}_{i}^{\dagger }\hat{c}_{j}^{{}}+V_{j,i}\hat{c}_{j}^{\dagger }\hat{c}%
_{i}^{{}}\right] \right\} .  \label{Hamil-sys}
\end{equation}%
Here, $\hat{c}_{i}^{\dagger }$ and $\hat{c}_{i}^{{}}$correspond to the
creation and anihilation fermionic operators acting on the vacuum $%
\left\vert 0\right\rangle $. \ Site energies are $E_{i}$ and hopping
amplitudes $V_{i,j}$ define the matrix Hamiltonian whose single particle
eigenstates are $\left\vert k\right\rangle =\sum_{i}u_{i,k}\hat{c}%
_{i}^{\dagger }\left\vert 0\right\rangle $ of energy $\varepsilon _{k}$
which are filled up to the Fermi energy, $\varepsilon _{F}$.

The decimation procedures, inspired in the renormalization group techniques
of statistical mechanics \cite{Kadanoff1983,Jose1982}, seek to recursively
reduce the number of degrees of freedom of a general $N\times N$ Hamiltonian
into another of lower rank, without altering the physical properties. The
basic idea can be captured by considering a system with $N=3$ states whose
secular equation is:%
\begin{equation}
\left[ 
\begin{array}{ccc}
\varepsilon -E_{1} & -V_{12} & -V_{13} \\ 
-V_{21} & \varepsilon -E_{2} & -V_{23} \\ 
-V_{31} & -V_{32} & \varepsilon -E_{3}%
\end{array}%
\right] \left( 
\begin{array}{c}
u_{1} \\ 
u_{2} \\ 
u_{3}%
\end{array}%
\right) =\left[ \varepsilon \mathbb{I}-\mathbb{H}_{S}\right] \overrightarrow{%
u}\equiv \overrightarrow{0}.
\end{equation}%
Quite often we are interested in the transfer of an excitation from an
initial state to another one, say 1 and 2. Thus, instead of diagonalizing
the matrix, we could isolate $u_{3}$ from the third row and use it to
eliminate $u_{3}$ in the first and the second equations. In this way, we
obtain a new set of equations where $u_{3}$ is \textit{decimated}:%
\begin{equation}
\left[ 
\begin{array}{cc}
\varepsilon -\overline{E}_{1} & -\overline{V}_{12} \\ 
-\overline{V}_{21} & \varepsilon -\overline{E}_{2}%
\end{array}%
\right] \left( 
\begin{array}{c}
u_{1} \\ 
u_{2}%
\end{array}%
\right)  \label{H_S2eff} \\
=[\varepsilon \mathbb{I}-\mathbb{H}_{\mathrm{eff.}}]\vec{u}=0.
\end{equation}%
The renormalized coefficients hide their non-linear dependence on the energy
variable $\varepsilon :$ 
\begin{equation}
\begin{array}{c}
\overline{E}_{1}=E_{1}+\Sigma _{1}(\varepsilon )=E_{1}+V_{13}\dfrac{1}{%
\varepsilon -E_{3}}V_{31}, \\ 
\overline{E}_{2}=E_{2}+\Sigma _{2}(\varepsilon )=E_{2}+V_{23}\dfrac{1}{%
\varepsilon -E_{3}}V_{32}, \\ 
\overline{V}_{12}=V_{12}+V_{13}\dfrac{1}{\varepsilon -E_{3}}V_{32}.%
\end{array}
\label{eq:decim-ex}
\end{equation}%
In this case, the terms $\Sigma _{j}(\varepsilon );j=1,2$ are the real
self-energies accounting for the energy shifts due to the coupling with the
eliminated state. Notice that as long as one conserves the analytical
dependence on $\varepsilon $ of $\Sigma _{j}$, the actual secular equation
is still cubic in $\varepsilon $ and provides the exact spectrum of the
whole system. This procedure can be performed systematically in a
Hamiltonian of any size $N\times N$ to end up with an effective Hamiltonian
of size one desires, in particular a $2\times 2$ one. The effective
interaction parameter $\overline{V}_{12}$, together with the self-energies $%
\Sigma _{j}$, accounts for transport through the whole sample. Their
dependence on $\varepsilon $ provides all the needed information on the
steady state transport as well as on quantum dynamics. \cite{Levstein1990}
In practice, it is convenient to add an infinitesimal imaginary part, $-%
\mathrm{i}\eta $, to each energy $E_{j}\rightarrow E_{j}-\mathrm{i}\eta $.
Since a finite $\eta >0$ is equivalent to a decay process, it ensures that
one recovers the retarded time dependences of the observables through a well
defined Fourier transform.

The terminals connected to the system are described as semi-infinite leads
coupled to it. They are handled in a similar way as the system itself. The
idea is to eliminate all the internal degrees of freedom decimating them
progressively, renormalizing the states of the system which are directly
coupled to the external reservoirs. For further clarification we consider a
lead modeled as a semi-infinite one dimensional chain, 
\begin{equation}
\hat{H}_{L}=\sum\limits_{i=0}^{-\infty }\left\{ E_{i}\hat{c}_{i}^{\dagger }%
\hat{c}_{i}^{{}}-V\left[ \hat{c}_{i}^{\dagger }\hat{c}_{i-1}^{{}}+\hat{c}%
_{i-1}^{\dagger }\hat{c}_{i}^{{}}\right] \right\} ,
\end{equation}%
that yields a tridiagonal matrix of infinite dimension. The elements $E_{i}$%
's and $V$'s are now the diagonal and off-diagonal terms of a tridiagonal
matrix $\mathbb{H}_{L}.$ This lead is connected at the left of the system,
say, with site $1$: 
\begin{equation}
\hat{V}_{SL}=V_{L}\left[ \hat{c}_{1}^{\dagger }\hat{c}_{0}^{{}}+\hat{c}%
_{0}^{\dagger }\hat{c}_{1}^{{}}\right] .
\end{equation}%
Instead of dealing with the whole Hamiltonian 
\begin{equation}
\hat{H}=\hat{H}_{S}+\hat{H}_{L}+\hat{V}_{SL},
\end{equation}%
we perform the decimation procedure. It becomes particularly simple because
of the chain structure of the lead. The energy of the $i$-th site, is \
\textquotedblleft shifted\textquotedblright\ by the elimination of $(i-1)$%
-th site, which itself is shifted by sites at its left \cite%
{Pastawski-Medina}, with the self-energies resulting in a
continued-fraction: 
\begin{eqnarray}
\Sigma _{i} &=&V_{i,i-1}\dfrac{1}{\varepsilon -E_{i-1}-\Sigma _{i-1}}%
V_{i-1,i} \\
(i &=&0,-1,-2,...-\infty )  \notag
\end{eqnarray}%
In a perfect propagating channel: $V_{i,i-1}\equiv V$ and $E_{i}=E_{0}$, and
thus, $\Sigma _{i}=\Sigma _{i-1}\equiv \Sigma $, we arrive to the
self-consistent solution: 
\begin{eqnarray}  \label{Dyson}
&&\Sigma (\varepsilon )=\dfrac{V^{2}}{\varepsilon -E_{0}-\Sigma }=\Delta
(\varepsilon )-\mathrm{i}\Gamma (\varepsilon ).  \notag  \label{sigma-leads}
\\
&=&\dfrac{\varepsilon -E_{0}+\mathrm{i}\eta }{2}-\text{sgn}(\varepsilon
-E_{0})\sqrt{\left( \dfrac{\varepsilon -E_{0}+\mathrm{i}\eta }{2}\right)
^{2}-V^{2}},
\end{eqnarray}%
where the generalized square root \cite{SquareRoot} in the limit $\eta
\rightarrow 0^{+},$ yields the imaginary component of the self-energy for $%
\varepsilon $ within the band of allowed energies. It becomes real otherwise.

Thus, once the states in the left lead are fully decimated the energy of the
first site becomes 
\begin{eqnarray}
\widetilde{E}_{1}(\varepsilon ) &=&\overline{E}_{1}(\varepsilon )+\Sigma
_{L1}(\varepsilon ) \\
\text{with~~}\Sigma _{L1}(\varepsilon ) &=&\left( \frac{V_{L}}{V_{{}}}%
\right) ^{2}\Sigma (\varepsilon ) \\
&=&\Delta _{L1}(\varepsilon )-\mathrm{i}\Gamma _{L1}(\varepsilon ) \label{self_energy}
\end{eqnarray}%
As before, the real part $\Delta _{L1}(\varepsilon )$ indicates how the
unperturbed site energies are shifted by the leads. The important difference
with the simple decimation example discussed above is that, as a consequence
of the infinite nature of the lead, the self-energies may acquire a finite
imaginary component, $\Gamma _{L1}(\varepsilon ),$ even in the limit $\eta
\rightarrow 0^{+}$. It describes the rate at which coherent density
excitation in the system decays into the lead propagating states.

Note that, the imaginary part is roughly consistent with the exponential
decays of the survival probability predicted by the Fermi Golden Rule (FGR).
For instance, in a \textquotedblleft system\textquotedblright\ with a single
state $\left\vert 1\right\rangle $ interacting with a lead, the survival
probability at time $t$ after it has been placed in state $\left\vert
1\right\rangle $ is, 
\begin{eqnarray}
\left\vert \left\langle 1\right\vert \exp [-\mathrm{i}\hat{H}~t/\hbar
]\left\vert 1\right\rangle \theta (t)\right\vert ^{2} &\equiv &\left\vert 
\mathrm{i}\hbar G_{11}^{R}(t)\right\vert ^{2} \\
&\simeq &\exp [-2\Gamma _{L1}(E_{1})t/\hbar ],
\end{eqnarray}%
where we introduced the time dependent retarded Green's function, $%
G_{11}^{R}(t)$. However, we remember that the self-energies obtained above
have an explicit functional dependence on $\varepsilon .$ In consequence,
the actual decay can depart from this naive exponential approximation.
Indeed, a quantum decay should start quadratically as $1-\left( V_{L}t/\hbar
\right) ^{2}$ turning into an exponential at very short times. At very long
times the decay may even become a non-monotonous. \cite{Rufeil2006} In
practice, we will stay in the exponential approximation by neglecting the
dependence on $\varepsilon $ unless it is close to a band edge.

\begin{figure*}[tbh]
\begin{center}
\includegraphics[width=5.5in]{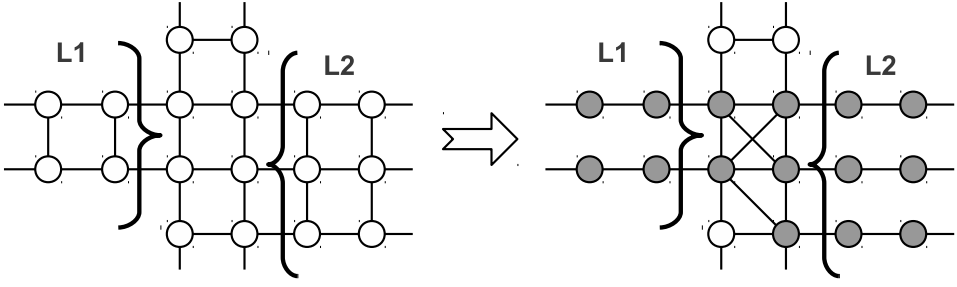}
\end{center}
\caption{Diagrammatic representation of an unitary transformation of the
system to a basis in which leads are independent. Here, dots represent
diagonal elements of the Hamiltonian in a site basis and lines non-diagonal
ones.}
\label{Gr:leads-indep}
\end{figure*}

For the sake of simplicity, we may idealize the terminal leads as quasi 1-D
wires. As waveguides, they can be described in terms of open channels at the
Fermi energy or propagating modes. Thus, we chose a basis for the system's
Hamiltonian in which each independent propagation mode $l$ of a lead is
connected to a single system's state. This might require a unitary
transformation to choose a system's basis that matches the propagating modes
of leads (see Fig. \ref{Gr:leads-indep}). There is no restriction to the
converse: i.e. each \textquotedblleft site\textquotedblright\ can be coupled
to different quantum channels. Since the leads can be represented by
homogeneous infinite tight-binding chains, their decimation is just the
procedure implemented above with the appropriate $V$'s and $E$'s describing
each mode $l$. 

The observation of DP was that any \textquotedblleft
local\textquotedblright\ electronic state weakly coupled to a huge number of
environmental degrees of freedom should decay from its initial decoupled
state according to the FGR. This would require a restitution or re-injection
of any escaping particle. Thus the DP model treats these decoherent
scattering channels sources as on-site fictitious voltage probes. Much as it
occurs with real voltmeters, local current conservation on each scattering
channels must be imposed. This ensures that each electron with definite
energy that escapes from a state towards a fictitious probe, is balanced by
an electron \textit{with the same energy} re-injected into the same state.
In the DP model, these decoherent channels are described by local
corrections to site energies of the sample, on the same footing as the real
channels: 
\begin{equation}
\hat{\Sigma}_{\phi i}=-\mathrm{i}\Gamma _{\phi i}\hat{c}_{i}^{\dagger }\hat{c%
}_{i}^{{}}.  \label{sigma-decoher}
\end{equation}%
Here, $\Gamma _{\phi i}$ represents an energy uncertainty associated with
the interaction process $\phi $ that mixes the local electron state $i$ with
environmental degrees of freedom. This introduces a decay of the state $i$
that can be described by the FGR. Notice that the state $i$ does not
necessarily represent a local basis, but it could be a channel mode or a
momentum basis state as well. The energy uncertainties due to decoherent
processes can be estimated for each specific process, \cite{PAni-CBP} and
may not necessarily be the same for every state $i$. Accordingly, each
\textquotedblleft site\textquotedblright\ $i$ may be subject to different
decay processes $\alpha $: those associated with real leads, $\alpha ={l,}$
and those related to decoherent processes (or fictitious probes), $\alpha ={%
\phi }$. The resulting effective Hamiltonian, $\hat{H}_{\mathrm{eff.}}$,
that includes the real and fictitious probes, is non-Hermitian \cite%
{Rotter2009}: 
\begin{equation}
\hat{H}_{\mathrm{eff.}}=(\hat{H}_{S}-\mathrm{i}\eta \hat{I}%
)+\sum\limits_{\alpha }\sum\limits_{i=1}^{N}\hat{\Sigma}_{\alpha i}.
\label{Hamil-efectivo}
\end{equation}%
Here, $\mathrm{Im}\hat{\Sigma}_{\alpha i}\neq 0$ only for those sites $i$
subject to decoherent processes ($\alpha =\phi $) or escapes to the leads ($%
\alpha =l$). Trivially, if the full imaginary part correction were
homogeneous (the same value for each state $i$), it just shifts the
eigenenergies into the complex plane. In contrast, inhomogeneous corrections
might produce spectral bifuctations that result in a quantum dynamical phase
transition. \cite{Dente2008}

In transport problems, most of the information on system dynamics is
distilled into the retarded and advanced Green functions. More practical
expressions are obtained using its Fourier transform into the energy
variable $\varepsilon $, from the effective Hamiltonian given by Eq. \ref%
{Hamil-efectivo}. In matrix representation: 
\begin{equation}
\mathbb{G}^{R}(\varepsilon )=\left[ \varepsilon \mathbb{I}-\mathbb{H}_{%
\mathrm{eff}.}\right] ^{-1}=\mathbb{G}^{A\dagger }(\varepsilon )
\label{eq:green-def}
\end{equation}%
These Green's functions contain all the information of the quantum system
coupled to the leads and environment and constitute the kernel to move into
the non-equilibrium problem. Also, diagonal elements provide the
\textquotedblleft local\textquotedblright\ density of states%
\begin{equation}
N_{i}(\varepsilon )=-\frac{1}{\pi }\mathrm{Im}G_{i,i}^{R}(\varepsilon )=-%
\frac{1}{2\pi \mathrm{i}}\left[ G_{i,i}^{R}(\varepsilon
)-G_{i,i}^{A}(\varepsilon )\right]
\end{equation}%
In particular, the transmission amplitudes of electronic excitations\
between the channels identified with process $\alpha $ at site $i$ and
process $\beta $ at site $j$ can be evaluated from the generalized form of
Fisher-Lee formula \cite{Pastawski-Medina}:%
\begin{equation}
t_{\alpha i,\beta j}(\varepsilon )=\mathrm{i}2~\sqrt{\Gamma _{\beta
j}^{{}}(\varepsilon )}~G_{j,i}^{R}(\varepsilon )~\sqrt{\Gamma _{\alpha
i}^{{}}(\varepsilon )}
\end{equation}%
and the transmission probabilities are given by:%
\begin{eqnarray}
T_{\alpha i,\beta j}(\varepsilon ) &=&\left\vert t_{\alpha i,\beta
j}(\varepsilon )\right\vert ^{2}\text{~~~~}(\alpha i\neq \beta j)  \notag \\
&=&4\Gamma _{\beta j}(\varepsilon )G_{j,i}^{R}(\varepsilon )\Gamma _{\alpha
i}(\varepsilon )G_{i,j}^{A}(\varepsilon )  \label{Fisher-Lee-generalizada}
\end{eqnarray}%
where ${\Gamma }_{\alpha i}=\mathrm{i}(\Sigma _{\alpha ,i}^{R}-\Sigma
_{\alpha ,i}^{A})/2$ is proportional the escape rate at site $i$ due to a
process $\alpha $.

\section{\label{sec:DP}Two-terminal D'Amato-Pastawski Model.}

Retarded and advanced Green's functions and the transmission probabilities
associated with them contain the basic quantum dynamics. In order to
describe the non-equilibrium properties of a system, one has to evaluate the
density matrix or simply the diagonal terms of non-equilibrium density
functions, 
\begin{equation}
G_{j,j}^{<}(\varepsilon )=\mathrm{i}2\pi N_{j}(\varepsilon )\mathrm{f}%
_{j}(\varepsilon ).  \label{eq:Keldyshdensitydef}
\end{equation}%
These, in turn, are determined by the boundary conditions imposed by the
external reservoirs $\beta j$ that act as a source or drain of particles.
Their occupation is described by a non-equilibrium distribution function
approximated by a shifted Fermi distribution $\mathrm{f}_{\beta
j}(\varepsilon )=1/(\exp [\left( \varepsilon -\varepsilon _{F}-\delta \mu
_{\beta j}\right) /k_{B}T])$. In the Quantum Fields formalism, the $G_{\phi
j,\phi j}^{<}(\varepsilon )$ Green's functions result from the quantum
evolution in presence of the boundary conditions. In the time independent
case, energy is conserved, and the non-equilibrium density function takes
the form, 
\begin{equation}
G_{j,k}^{<}(\varepsilon )=2\mathrm{i}\sum\limits_{\alpha
i}G_{j,i}^{R}(\varepsilon )\Gamma _{\alpha i}(\varepsilon )\mathrm{f}%
_{\alpha i}(\varepsilon )G_{i,k}^{A}(\varepsilon ),
\label{eq:Keldyshdensity-integral}
\end{equation}%
i.e. densities and correlations inside the system result from the
occupations $\mathrm{f}_{\beta i}(\varepsilon )$ imposed by the
experimentalist at the current terminals and the environment at the\
\textquotedblleft fictitious\textquotedblright\ probes. The equilibrium
density function $G_{j,j}^{(0)<}(\varepsilon )$ results when $\delta \mu
_{\beta j}\equiv 0$ for all $\beta j$. The actual observables are evaluated
from this non-equilibrium density function. The change respect to the
equilibrium in the local density can be expressed in terms of the above
boundary conditions as \cite{GLBE2}: 
\begin{eqnarray}
\delta \rho _{j}^{{}} &=&-\frac{\mathrm{i}}{2\pi }\int \left[
G_{j,j}^{<}-G_{j,j}^{(0)<}\right] \mathrm{d}\varepsilon \\
&\simeq &N_{j}(\varepsilon _{F})\delta \mu _{j},  \notag
\end{eqnarray}%
while the currents between sites $i$ and $j$ are given by%
\begin{equation}
I_{i,j}=\int \left[ V_{i,j}G_{j,i}^{<}-V_{j,i}G_{i,j}^{<}\right] \mathrm{d}%
\varepsilon .
\end{equation}%
These integral expressions of the observables, expressed in the linear
response approximation of small biases $e\mathtt{V}_{L}=\mu
_{Li}-\varepsilon _{F}\ll \varepsilon _{F}$, become the Generalized
Landauer-B\"{u}ttiker equations that describe the balance of electronic
current. These are no other than the Kirchhoff laws expressed in terms of
the generalized Landauer's conductances, given by the Fisher-Lee formulas of
Eq. \ref{Fisher-Lee-generalizada}. Because of the linear approximation these
transmittances are evaluated at the Fermi energy, and now become: 
\begin{equation}
I_{\alpha i}=\frac{e}{h}\underset{\text{processes}}{\sum\limits_{\beta
=L,\phi }}\underset{}{}\underset{\text{sites}}{\sum\limits_{j=1(\alpha i\neq
\beta j)}^{N}}\left( T_{\alpha i,\beta j}\delta \mu _{\beta j}-T_{\beta
j,\alpha i}\delta \mu _{\alpha i}\right)  \label{eq:Kirchhoff1}
\end{equation}%
where the quantities $\delta \mu _{\alpha i}=\mu _{\alpha i}-\varepsilon
_{F},$ are the chemical potentials of the electron reservoirs, at state $i$
for a process $\alpha $.

The requirement in the DP model that no net current flows through the
decoherent channels imposes 
\begin{equation}
0\equiv I_{\phi i}.  \label{eq:CurrentConservation}
\end{equation}%
These equations imply the self-consistent determination of the internal
non-equilibrium chemical potentials $\delta \mu _{\phi i}.$ Thus, we are
faced to a linear problem. Once again, its solution can be laid as a
decimation procedure, as we did to obtain the effective Hamiltonian.

Consider the case where two real leads are connected to the sites $1$ and $N$
of the system (thus identified as channels $\ell 1$ and $\ell N$), and a 
\textit{single} decoherent process $\phi k$ is connected to the state $k$.
Thus, charge conservation implies: 
\begin{equation}
0=T_{\phi k,\ell 1}\delta \mu _{\ell 1}+T_{\phi k,\ell N}\delta \mu _{\ell
N}-(T_{\ell 1,\phi k}+T_{\ell N,\phi k})\delta \mu _{\phi k},
\label{eq:Decim-T_first}
\end{equation}%
which can be rewritten as: 
\begin{equation}
\delta \mu _{\phi k}=\frac{T_{\phi k,\ell N}}{(T_{\ell 1,\phi k}+T_{\ell
N,\phi k})}\delta \mu _{LN}+\frac{T_{\phi k,\ell 1}}{(T_{\ell 1,\phi
k}+T_{\ell N,\phi k})}\delta \mu _{\ell 1}
\end{equation}%
Using this relation for the current on real channels we obtain: 
\begin{equation}
I_{\ell N}=-I_{\ell 1}=\frac{e}{h}\tilde{T}_{\ell N,\ell 1}(\delta \mu
_{\ell N}-\delta \mu _{\ell 1}),  \label{eq:currentDP}
\end{equation}%
where $\tilde{T}_{\ell N,\ell 1}$ represent the \textquotedblleft
effective\textquotedblright\ transmission between leads $\ell 1$ and $\ell N$
after the decimation of the incoherent channel associated with $\phi k$,
given by: 
\begin{equation}
\tilde{T}_{\ell N,\ell 1}=T_{\ell N,\ell 1}+T_{\ell N,\phi k}\frac{1}{%
(T_{\ell 1,\phi k}+T_{\ell N,\phi k})}T_{\phi k,\ell 1}.  \label{eq:Decim-T}
\end{equation}%
Note that the zero current constrain at the decoherent channels allows us to
pile up (i.e. decimate) those processes into an incoherent contribution to
the total transmission. This is the reason why Eq. \ref%
{eq:CurrentConservation} is the key factor in the computation of the total
transmission. At this point one recognizes the analogy of the second term on
the right-hand side of Eq. \ref{eq:Decim-T} with the effective interaction
shown in Eq. \ref{eq:decim-ex}. This analogy will be used in the following
section to develop a simple matrix solution for the total decoherent
transmission in a multi-terminal setup. In the case of two current probes,
identifying the index label $L=\ell 1$ and $R=\ell N$ for the leads, and $%
\phi k=k$ for the decoherence probes, one has that the total transmission
probability is given by:\cite{Damato-Pastawski} 
\begin{equation}
\tilde{T}_{L,R}=T_{L,R}+\sum\limits_{i,j}T_{R,i}\left[ \mathbb{W}^{-1}\right]
_{i,j}T_{j,L}.  \label{eq:Teff_DP}
\end{equation}%
The elements of the matrix $\mathbb{W}$ are: 
\begin{equation}
W_{ij}=-T_{ij}+\left( \sum\limits_{j=L,i,R}T_{ij}\right) \delta _{ij}.
\end{equation}

Eqs. \ref{eq:currentDP} and \ref{eq:Teff_DP} provide the decoherent current
and the effective transmission of DP model for two-terminal setups. However,
they need to be reformulated to deal with a multi-terminal setup as when
there are more than two externally controlled chemical potentials or when
one requires to discriminate among different processes that contribute to
the current.

\section{\label{sec:Computational}Multi-Terminal D'Amato-Pastawski Model}

The two-probe Landauer conductance requires the computation of a single
element of the Green's function matrix: that connecting sites where the
leads are attached. In a 1-D case, this is $G_{1N}$ (where $N$ is the number
of sites of the system) and can be calculated through a decimation procedure.%
\cite{Levstein1990} While this can be readily generalized to deal with
finite systems of any dimension, not all formulations result numerically
stable in presence of strong disorder or band gaps.\cite{PastawskiSlutzky}
We will present a particular algorithm that is stable in such conditions.
The method is applicable to block tridiagonal Hamiltonians. These are very
common in many physicaly relevant situations, specifically when interactions
are truncated, or when the Hamiltonian matrix presents some form of banded
structure.

The DP model requires the computation of the transmittances among all
possible pairs of fictitious and physical probes, roughly $M(M-1)/2$, where $%
M~(\leq N)$ is the number of phase-breaking scattering channels. Also the
computation of the effective transmission requires the inversion of $\mathbb{%
W}$, a $M\times M$ matrix, as expressed in Eq. \ref{eq:Teff_DP}. It is our
purpose to extend the scheme of the DP model to account for decoherence in
quantum transport problems that involves many terminals. We seek for a
decoherent transmission analogous to Eq. \ref{eq:Teff_DP} for each pair of
physical leads. Thus, the computational approach to the DP model would
require an efficient matrix inversion algorithm.

In the next subsection, we present a computational procedure that, being
based on decimation schemes, preserves the physical meaning of matrix
inversions. This may allow one to take advantage of system's symmetries as
they can usually be expressed as relations between $\mathbb{G}$'s elements.

\subsection{Green's Function and recursive algorithms.}

In order to obtain the Green's functions of Eq. \ref{eq:green-def}, a matrix
inversion is needed. The \textit{matrix continued fractions} \cite%
{Butler1973,MCF-Pastawski} scheme offers a decimative approach well suited
to perform this task. This procedure can be constructed recalling the well
known $2\times 2$ block matrix inversion, 
\begin{widetext}
\begin{equation}
\left[ 
\begin{array}{cc}
\mathbb{A} & \mathbb{B} \\ 
\mathbb{C} & \mathbb{D}%
\end{array}%
\right] ^{-1}=\left[ 
\begin{array}{cc}
(\mathbb{A}-\mathbb{BD}^{-1}\mathbb{C})^{-1} & -\mathbb{A}^{-1}\mathbb{B}(%
\mathbb{D}-\mathbb{CA}^{-1}\mathbb{B})^{-1} \\ 
-\mathbb{D}^{-1}\mathbb{C}(\mathbb{A}-\mathbb{BD}^{-1}\mathbb{C})^{-1} & (%
\mathbb{D}-\mathbb{CA}^{-1}\mathbb{B})^{-1}%
\end{array}%
\right] ,  \label{eq:BlockInvert}
\end{equation}%
\end{widetext}where $\mathbb{A}$, $\mathbb{B}$, $\mathbb{C}$ and $\mathbb{D}$
are arbitrary size subdivisions of the original matrix.

Let's assume that we have an effective Hamiltonian, $\hat{H}_{\mathrm{eff.}}$
which has block tridiagonal structure. We start \textquotedblleft
partitioning\textquotedblright\ the basis states in two portions: a cluster
labeled as $1$ that contains the first block, and the cluster of remaining
states of the system which we label as $B$. Thus, the Green's function
matrix in Eq. \ref{eq:green-def} is subdivided into four blocks, $%
(\varepsilon \mathbb{I}-\mathbb{E}_{1})$,$(\varepsilon \mathbb{I}-\mathbb{E}%
_{B})$,$-\mathbb{V}_{1B}$, and $-\mathbb{V}_{B1}$ of dimensions $N_{1}\times
N_{1}$, $N_{B}\times N_{B}$, $N_{1}\times N_{B}$ and $N_{B}\times N_{1}$
respectively. Thus, 
\begin{equation}
\mathbb{G}(\varepsilon )=\left[ 
\begin{array}{cc}
\mathbb{G}_{11} & \mathbb{G}_{1B} \\ 
\mathbb{G}_{B1} & \mathbb{G}_{BB}%
\end{array}%
\right] =\left[ 
\begin{array}{cc}
\varepsilon \mathbb{I}-\mathbb{E}_{1} & -\mathbb{V}_{1B} \\ 
-\mathbb{V}_{B1} & \varepsilon \mathbb{I}-\mathbb{E}_{B}%
\end{array}%
\right] ^{-1}.  \label{eq:Green-block}
\end{equation}%
Here, it is important to recall that the effective Hamiltonian $\hat{H}%
_{eff.}$ already includes all corrections due to fictitious and real probes,
by virtue of Eq. \ref{Hamil-efectivo}. In this way, the block with energies
and interactions, denoted here by $\mathbb{E}_{i}$, contain the
self-energies that account for the openness of the system, and may be
complex numbers. Combining Eq. \ref{eq:BlockInvert} and Eq. \ref%
{eq:Green-block} is easy to show that, 
\begin{equation}
\begin{array}{c}
\mathbb{G}_{11}=\left( \varepsilon \mathbb{I}-{\mathbb{E}}_{1}-\mathbf{%
\Sigma }_{1}^{(B)}\right) ^{-1}=\left( \varepsilon \mathbb{I}-\tilde{\mathbb{%
E}}_{1}\right) ^{-1}, \\ 
\mathbb{G}_{BB}=\left( \varepsilon \mathbb{I}-{\mathbb{E}}_{B}-\mathbf{%
\Sigma }_{B}^{(1)}\right) ^{-1}=\left( \varepsilon \mathbb{I}-\tilde{\mathbb{%
E}}_{B}\right) ^{-1}, \\ 
\mathbb{G}_{1B}=\mathbb{G}_{11}\mathbb{V}_{1B}(\varepsilon \mathbb{I}-%
\mathbb{E}_{B})^{-1}=\mathbb{G}_{11}\left[ \mathbf{\Sigma }_{1}^{(B)}\mathbb{%
V}_{B1}^{-1}\right] ,\text{ and} \\ 
\mathbb{G}_{B1}=\mathbb{G}_{BB}\mathbb{V}_{B1}(\varepsilon \mathbb{I}-%
\mathbb{E}_{1})^{-1}=\mathbb{G}_{BB}\left[ \mathbf{\Sigma }_{B}^{(1)}\mathbb{%
V}_{1B}^{-1}\right] .%
\end{array}
\label{eq:Green-blocks2}
\end{equation}%
Here, the similarity with Eq. \ref{eq:decim-ex} allows us to define the
block self energies, $\mathbf{\Sigma }$'s, which in this simple $2\times 2$
block scheme, are given by: 
\begin{equation}
\begin{array}{c}
\left[ \mathbf{\Sigma }_{1}^{(B)}\mathbb{V}_{B1}^{-1}\right] =\left[ \mathbb{%
V}_{1B}(\varepsilon \mathbb{I}-\mathbb{E}_{B})^{-1}\right] , \\ 
\left[ \mathbf{\Sigma }_{B}^{(1)}\mathbb{V}_{1B}^{-1}\right] =\left[ \mathbb{%
V}_{B1}(\varepsilon \mathbb{I}-\mathbb{E}_{1})^{-1}\right] .%
\end{array}
\label{eq:Sigmas1}
\end{equation}%
Notice, that in the expressions of Eqs. \ref{eq:Green-blocks2} and \ref%
{eq:Sigmas1}, the inverse of the hopping matrix must cancel with the hopping
that enters in the self-energies definition. Since the hoppings may be
non-square matrices, this definition is crucial to avoid its inversion.
Considering the bracket factors $\left[ \mathbf{\Sigma }\mathbb{V}_{{}}^{-1}%
\right] $ as a single object ensures stability of the recurrence procedure.
The decimation of the degrees of freedom associated with the portion $B$ of
the effective Hamiltonian is implied in Eq. \ref{eq:Green-blocks2}, where: 
\begin{equation}
\tilde{\mathbb{E}}_{1}=\mathbb{E}_{1}+\mathbf{\Sigma }_{1}^{(B)}=\mathbb{E}%
_{1}+\left[ \mathbb{V}_{1B}(\varepsilon \mathbb{I}-\mathbb{E}_{B})^{-1}%
\right] \mathbb{V}_{B1}.  \label{eq:decimation_block1+R}
\end{equation}%
Likewise, the decimation of block $1$ into $B$ gives the effective block: 
\begin{equation}
\tilde{\mathbb{E}}_{B}=\mathbb{E}_{B}+\mathbf{\Sigma }_{B}^{(1)}=\mathbb{E}%
_{B}+\left[ \mathbb{V}_{B1}(\varepsilon \mathbb{I}-\mathbb{E}_{1})^{-1}%
\right] \mathbb{V}_{1B}.  \label{eq:decimation_blockR+1}
\end{equation}%
Note that with the adopted notation for the self energies, $\Sigma
_{i}^{(j)} $ is the correction to block site $i$ when all block sites
between $i$ and $j $ (with $j$ included) are decimated. Therefore the
supra-index in parentheses indicate the subspace that has been decimated.

Since we are dealing with tridiagonal block matrices, we may resort to a
further partition for the matrix inversion involved in Eq. \ref%
{eq:decimation_block1+R}. i.e. the block $B$ describes states that can be
subdivided into two clusters where the first one, labeled 2, corresponds to
the first tridiagonal block from $(\varepsilon \mathbb{I}-\mathbb{E}_{B})$.
The other block $B^{\prime }$ now satisfies $\mathbb{V}_{1B^{\prime }}\equiv 
\mathbb{O}$. Then, we have%
\begin{equation}
\mathbb{G}(\varepsilon )=\left[ 
\begin{array}{c|cc}
\varepsilon \mathbb{I}-\mathbb{E}_{1} & -\mathbb{V}_{12} & \mathbb{O} \\ 
\hline
-\mathbb{V}_{21} & \varepsilon \mathbb{I}-\mathbb{E}_{2} & -\mathbb{V}%
_{2B^{\prime }} \\ 
\mathbb{O} & -\mathbb{V}_{B^{\prime }2} & \varepsilon \mathbb{I}-\mathbb{E}%
_{B^{\prime }}%
\end{array}%
\right] ^{-1}.
\end{equation}%
Again, we can also decimate the degrees of freedom associated with block $2$%
, taking 
\begin{equation}
\begin{array}{cc}
\tilde{\mathbb{E}}_{1}=\mathbb{E}_{1}+\mathbf{\Sigma }_{1}^{(2)}, & \tilde{%
\mathbb{E}}_{B^{\prime }}=\mathbb{E}_{B^{\prime }}+\mathbf{\Sigma }%
_{B^{\prime }}^{(2)} \\ 
\multicolumn{2}{c}{\tilde{\mathbb{V}}_{1B^{\prime }}=\mathbb{V}%
_{12}(\varepsilon \mathbb{I}-\mathbb{E}_{2})^{-1}\mathbb{V}_{2B^{\prime }}}%
\end{array}%
\end{equation}%
which leads to an effective equation analogous to Eq. \ref{eq:Green-block},
in terms of the new effective block sites: 
\begin{equation}
\left[ 
\begin{array}{cc}
\mathbb{G}_{11} & \mathbb{G}_{1B^{\prime }} \\ 
\mathbb{G}_{B^{\prime }1} & \mathbb{G}_{B^{\prime }B^{\prime }}%
\end{array}%
\right] =\left[ 
\begin{array}{cc}
\varepsilon \mathbb{I}-\tilde{\mathbb{E}}_{1} & -\tilde{\mathbb{V}}%
_{1B^{\prime }} \\ 
-\tilde{\mathbb{V}}_{B^{\prime }1} & \varepsilon \mathbb{I}-\tilde{\mathbb{E}%
}_{B^{\prime }}%
\end{array}%
\right] ^{-1}
\end{equation}%
Therefore, an expression analogous to Eq. \ref{eq:Green-blocks2} is
obtained: 
\begin{equation}
\begin{array}{c}
\mathbb{G}_{11}=\left( \varepsilon \mathbb{I}-\mathbb{E}_{1}-\mathbf{\Sigma }%
_{1}^{(B^{\prime })}\right) ^{-1} \\ 
\mathbb{G}_{B^{\prime }B^{\prime }}=\left( \varepsilon \mathbb{I}-\mathbb{E}%
_{B^{\prime }}-\mathbf{\Sigma }_{B^{\prime }}^{(1)}\right) ^{-1} \\ 
\mathbb{G}_{1B^{\prime }}=\mathbb{G}_{11}\tilde{\mathbb{V}}_{1B^{\prime
}}(\varepsilon \mathbb{I}-\tilde{\mathbb{E}}_{B^{\prime }})^{-1} \\ 
\mathbb{G}_{B^{\prime }1}=\mathbb{G}_{B^{\prime }B^{\prime }}\tilde{\mathbb{V%
}}_{B^{\prime }1}(\varepsilon \mathbb{I}-\tilde{\mathbb{E}}_{1})^{-1}%
\end{array}
\label{eq:Green-blocks3}
\end{equation}%
where the diagonal blocks of the Green's function matrix involve%
\begin{equation}
\begin{array}{c}
\mathbf{\Sigma }_{1}^{(B^{\prime })}=\left[ \mathbb{V}_{12}(\varepsilon 
\mathbb{I}-\mathbb{E}_{2}-\mathbf{\Sigma }_{2}^{(B^{\prime })})^{-1}\right] 
\mathbb{V}_{12} \\ 
\mathbf{\Sigma }_{B^{\prime }}^{(1)}=\left[ \mathbb{V}_{B^{\prime
}2}(\varepsilon \mathbb{I}-\mathbb{E}_{2}-\mathbf{\Sigma }_{2}^{(1)})^{-1}%
\right] \mathbb{V}_{2B^{\prime }}%
\end{array}
\label{eq:Sigmas_border}
\end{equation}%
Note that in the self-energies of Eq. \ref{eq:Sigmas_border}, the decimated
space (denoted by the supra-index) always includes one of the border blocks
(in this case, $1$ or $B$). However, as shown hereafter, the non-diagonal
terms can also be written in terms of the block self-energies $\mathbf{%
\Sigma }^{(1)}$'s and $\mathbf{\Sigma }^{(B^{\prime })}$'s:%
\begin{equation}
\begin{array}{c}
\mathbb{G}_{1B^{\prime }}=\mathbb{G}_{11}[\mathbf{\Sigma }_{1}^{(B^{\prime
})}\mathbb{V}_{12}^{-1}][\mathbf{\Sigma }_{2}^{(B^{\prime })}\mathbb{V}%
_{B^{\prime }2}^{-1}], \\ 
\mathbb{G}_{B^{\prime }1}=\mathbb{G}_{B^{\prime }B^{\prime }}[\mathbf{\Sigma 
}_{B^{\prime }}^{(1)}\mathbb{V}_{1B^{\prime }}^{-1}][\mathbf{\Sigma }%
_{2}^{(1)}\mathbb{V}_{12}^{-1}].%
\end{array}
\label{eq:G1Bprime}
\end{equation}%
Both expressions are crucial to visualize the seed of our recursive
procedure.

\begin{figure*}[tbph]
\begin{center}
\includegraphics[width=5.5in]{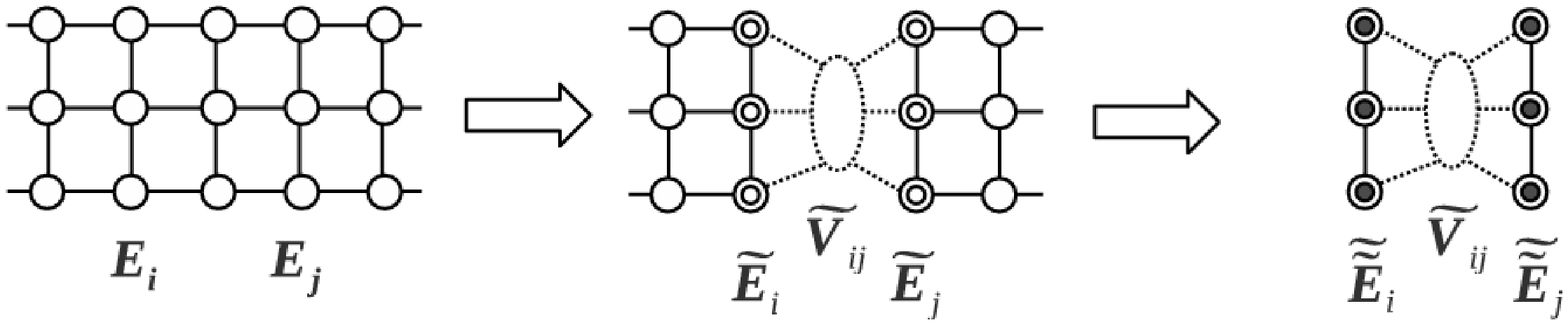}
\end{center}
\caption{Decimation scheme for the calculation of the elements of the
Green's function matrix.}
\label{Gr:Decimation}
\end{figure*}

The generalization by further partition into an arbitrary number of clusters
is straightforward. The Green's functions are expressed as a product of
non-singular self-energy blocks that are calculated recursively.
Independently of how the effective Hamiltonian is subdivided, if there are $%
N $ blocks of arbitrary size and the entire system is decimated into the $i$%
-th and $j$-th block, we have simply as matrix continued fractions: \cite%
{MCF-Pastawski} 
\begin{equation}
\begin{array}{c}
\mathbf{\Sigma }_{i}^{(j)}=\left[ \mathbb{V}_{i,i+1}\left( \varepsilon 
\mathbb{I}-\mathbb{E}_{i+1}-\mathbf{\Sigma }_{i+1}^{(j)}\right) ^{-1}\right] 
\mathbb{V}_{i+1,i} \\ 
\mathbf{\Sigma }_{j}^{(i)}=\left[ \mathbb{V}_{j,j-1}\left( \varepsilon 
\mathbb{I}-\mathbb{E}_{j-1}-\mathbf{\Sigma }_{j-1}^{(i)}\right) ^{-1}\right] 
\mathbb{V}_{j-1,j}^{{}} \\ 
\text{for}~~~~j>i,%
\end{array}%
\end{equation}%
provided that the final structure preserves a block three-diagonal. We
recall that matrix inversions are further stabilized by the presence of the
imaginary site energies imposed by the real and fictitious probes (Eq. \ref%
{Hamil-efectivo}). In this way, the decimation of the entire system into the
arbitrary \textquotedblleft block\textquotedblright\ sites $i$ and $j$,
leads to the effective quantities 
\begin{equation}
\begin{array}{c}
\tilde{\mathbb{E}}_{i}=\mathbb{E}_{i}+\mathbf{\Sigma }_{i}^{(1)}+\mathbf{%
\Sigma }_{i}^{(j)} \\ 
\tilde{\mathbb{E}}_{j}=\mathbb{E}_{j}+\mathbf{\Sigma }_{j}^{(i)}+\mathbf{%
\Sigma }_{j}^{(N)} \\ 
\tilde{\mathbb{V}}_{i,j}=\tilde{\mathbb{V}}_{i,j-1}(\varepsilon \mathbb{I}-%
\mathbb{E}_{j}-\mathbf{\Sigma }_{j}^{(1)})^{-1}\mathbb{V}_{j-1,j}%
\end{array}%
\end{equation}%
\textit{which determine exactly} each $(i,j)$ element of the total Green's
function, 
\begin{equation}
\left[ 
\begin{array}{cc}
\mathbb{G}_{ii} & \mathbb{G}_{ij} \\ 
\mathbb{G}_{ij} & \mathbb{G}_{jj}%
\end{array}%
\right] =\left[ 
\begin{array}{cc}
\varepsilon \mathbb{I}-\tilde{\mathbb{E}}_{i} & -\tilde{\mathbb{V}}_{ij} \\ 
-\tilde{\mathbb{V}}_{ji} & \varepsilon \mathbb{I}-\tilde{\mathbb{E}}_{j}%
\end{array}%
\right] ^{-1}.
\end{equation}%
The last expression is similar to Eq. \ref{eq:Green-block}, and therefore we
have, 
\begin{equation}
\begin{array}{c}
\mathbb{G}_{ii}=\left[ (\varepsilon \mathbb{I}-{\mathbb{E}}_{i})-\mathbf{%
\Sigma }_{i}^{(1)}-\mathbf{\Sigma }_{i}^{(N)}\right] ^{-1}, \\ 
\mathbb{G}_{jj}=\left[ (\varepsilon \mathbb{I}-{\mathbb{E}}_{j})-\mathbf{%
\Sigma }_{j}^{(1)}-\mathbf{\Sigma }_{j}^{(N)}\right] ^{-1}, \\ 
\mathbb{G}_{ij}=\mathbb{G}_{ii}\left[ \tilde{\mathbb{V}}_{ij}(\varepsilon 
\mathbb{I}-\tilde{\mathbb{E}}_{j})^{-1}\right] , \\ 
\mathbb{G}_{ji}=\mathbb{G}_{jj}\left[ \tilde{\mathbb{V}}_{ji}(\varepsilon 
\mathbb{I}-\tilde{\mathbb{E}}_{i})^{-1}\right] .%
\end{array}
\label{eq:Green-blocksF}
\end{equation}%
This procedure is shown diagrammatically on Fig. \ref{Gr:Decimation}. 
Note that the diagonal elements are easily calculated evaluating $\mathcal{O(%
}N)$ energy corrections of the form $\Sigma _{i}^{(1)}$ and $\Sigma
_{i}^{(N)}$, where all the sites have been decimated into site $i$. Also, in
order to compute all the non diagonal elements of the Green's function
matrix in Eq. \ref{eq:Green-blocks3} and \ref{eq:Green-blocksF} we would
need to evaluate $\sim N^{2}$ energy corrections $\mathbf{\Sigma }_{i}^{(j)}$%
's. However, following the insight given in Eq. \ref{eq:G1Bprime}, for
tridiagonal block Hamiltonians, the non-diagonal block matrix elements of
the Green function can be obtained in terms of the diagonal ones, avoiding
the need of the evaluation of $\mathcal{O(}N^{2})$ terms $\Sigma _{i}^{(j)}$%
's. In this case, if the Hamiltonian matrix is subdivided in $N$ arbitrary
blocks, we have 
\begin{eqnarray}
&&\underset{\text{where}~i<j}{\mathbb{G}_{ij}=\mathbb{G}_{ii}\prod%
\limits_{k=i}^{j-1}\left[ \mathbf{\Sigma }_{k}^{(N)}\mathbb{V}_{k+1,k}^{-1}%
\right] },  \label{eq:thoulessR} \\
&&\underset{\text{where}~i<j}{\mathbb{G}_{ji}=\mathbb{G}_{jj}\prod%
\limits_{k=j}^{i+1}\left[ \mathbf{\Sigma }_{k}^{(1)}\mathbb{V}_{k-1,k}^{-1}%
\right] }.  \label{eq:thoulessL}
\end{eqnarray}%
Note that now it is not necessary to evaluate any extra $\mathbf{\Sigma }$
in order to calculate $\mathbb{G}_{ij}$ for $i\neq j$, because those
self-energies have been already calculated for the diagonal Green's
Functions matrix blocks, $\mathbb{G}_{ii}$. This implies that only $\mathcal{%
O}\left( N\right) $ self-energies are required for the calculation of the
whole Green's function. These equations can help to take advantage of
possible symmetries of the $\mathbb{V}$ and $\mathbf{\Sigma }$ matrices to
speed up even more the calculation of Green's Functions.

Although Eqs. \ref{eq:thoulessR}-\ref{eq:thoulessL} have been formally
written in terms of hopping matrix inverses, $\mathbb{V}^{-1}$, these
expressions are accurate even when the hopping matrices are singular. This
is because the hopping matrix inverse cancels out with the hopping in the $%
\Sigma $ definition, as it can be seen, for example, in Eq. \ref{eq:Sigmas1}%
. In most cases, $\mathbb{G}_{ij}^{R}=\mathbb{G}_{ji}^{R}$, and therefore
Eqs. \ref{eq:thoulessR} and \ref{eq:thoulessL} are equivalent. However, both
equations are needed in some cases of quantum pumping \cite{LuisMPQP} or in
the presence of magnetic fields. The origin of the extraordinary stability
of Eqs. \ref{eq:thoulessR}-\ref{eq:thoulessL} can be easily grasped
analytically by considering a linear chain with three sites and expressing
the self-energies in terms of continued fractions before applying Eq. \ref%
{eq:G1Bprime}. Explicitly,%
\begin{widetext}
\begin{equation}
G_{1,3}(\varepsilon )=\cfrac{1}{\varepsilon -E_{1}-V_{12}\cfrac{1}{%
\varepsilon -E_{2}-V_{23}\cfrac{1}{\varepsilon -E_{3}}V_{32}}V_{21}}V_{12}%
\cfrac{1}{\varepsilon -E_{2}-V_{23}\cfrac{1}{\varepsilon -E_{3}}V_{32}}V_{23}%
\cfrac{1}{\varepsilon -E_{3}}.  \label{Eq-G13}
\end{equation}
\end{widetext}Here, we clearly see that the divergences in the last factor
are exactly canceled by the zeros of the second one, while the singularities
in this one, are canceled by the zeros in the first factor. This equation
holds when the elements $E$'s and $V$'s are replaced by matrices, with $%
\mathbb{V}_{n,n+1}$'s mixing subspaces $\mathbb{E}_{n}$ and $\mathbb{E}%
_{n+1} $ of different dimensions. In general, the divergences in $\left[ 
\mathbf{\Sigma }_{k}^{(N)}\mathbb{V}_{k-1,k}^{-1}\right] $ are compensated
by the zeros of the previous term, i.e. $\left[ \mathbf{\Sigma }_{k-1}^{(N)}%
\mathbb{V}_{k,k-1}^{-1}\right] $. Furthermore, the regularization of poles
and divergencies imposed by decoherent processes (see below) ensure the
numerical precision of this cancellation.

\subsection{Physical Observables in the Multi-Terminal D'Amato-Pastawski
model}

The application of the DP model for multi-terminal devices requires a
generalization of Eq. \ref{eq:Teff_DP}. To obtain the total transmission on
each terminal, we can take advantage of the decimation procedures discussed
above. Eq. \ref{eq:Kirchhoff1} is easily rearranged in terms of the
transmissivity $(1-R_{\alpha i})$ from each channel $\alpha i$. \cite%
{Pastawski-Medina} For process $\alpha $ at site $i$, one defines 
\begin{eqnarray}
\left\vert t_{\alpha i,\alpha i}\right\vert ^{2}+(1-R_{\alpha i})
&=&\left\vert t_{\alpha i,\alpha i}\right\vert ^{2}\underset{\left( \beta
j\neq \alpha i\right) }{+\sum_{\beta ,j}}T_{\beta j,\alpha i}  \label{Eq:1-R}
\\
&=&(1/g_{\alpha ,i})=4\pi N_{i}\Gamma _{\alpha i},  \label{Eq:g-T}
\end{eqnarray}%
where $N_{i}$ is the density of states at the site $i$. The Fisher-Lee
formula is extended by defining a \textquotedblleft
self-transmission\textquotedblright\ $\left\vert t_{\alpha i,\alpha
i}\right\vert ^{2}$ that is not a transmittance in the standard sense, and
certainly it is not the diagonal term $T_{\alpha i,\alpha i}\equiv R_{\alpha
i}-1$. However, it is required to obtain the sum of Eq. \ref{Eq:g-T} as the
product of the local density of states and the decay rate. It describes all
the electrons that, at a certain instant, are leaving the $\alpha i^{\mathrm{%
th}}$ reservoir to eventually return after wandering around. The inclusion
of this term is important because it contributes to define $(1/g_{\alpha
,i}) $, which plays a central role in a Keldysh perturbative expansion \cite%
{GLBE1,Pastawski-Medina} and in a time dependent formulation of transport. 
\cite{GLBE2}

Therefore, in a steady state calculation is enough to express eq. \ref%
{eq:Kirchhoff1} as: 
\begin{equation}
I_{\alpha i}=\frac{\left\vert e\right\vert }{h}\left[ (R_{\alpha i}-1)\delta
\mu _{\alpha i}+\sum\limits_{\beta =L,\phi }\underset{\alpha i\neq \beta j}{%
\sum\limits_{j=1}^{N}}T_{\alpha i,\beta j}\delta \mu _{\beta ,j}\right] .
\label{Eq:Kirchhoff-current}
\end{equation}%
It can be arranged in a compact matrix notation, separating the processes
associated with the leads from the decoherent ones. The actual currents at
the leads are arranged in the vector $\overrightarrow{I}_{\lambda }$ while
the vanishing currents at the decoherent channels, in $\overrightarrow{I}%
_{\phi }\equiv \overrightarrow{0}$. Thus, 
\begin{equation}
\left( 
\begin{array}{c}
\overrightarrow{I}_{\lambda } \\ 
\overrightarrow{0}%
\end{array}%
\right) =\frac{\left\vert e\right\vert }{h}\mathbb{T}\left( 
\begin{array}{c}
\delta \overrightarrow{\mu _{\lambda }} \\ 
\delta \overrightarrow{\mu _{\phi }}%
\end{array}%
\right) .  \label{eq:CurrentsMatrix}
\end{equation}%
Here, the non-diagonal elements of $\mathbb{T}$ are transmission
probabilities and thus, they are definite positive. In contrast, the
diagonal elements are negative. Thus, a sum over any column or row cancels
out. This matrix can also be subdivided in the same block structure: 
\begin{equation}
\mathbb{T}=\left[ 
\begin{array}{cc}
\mathbb{T}_{\lambda \lambda } & \mathbb{T}_{\lambda \phi } \\ 
\mathbb{T}_{\phi \lambda } & \mathbb{T}_{\phi \phi }%
\end{array}%
\right] .
\end{equation}%
This notation stress that $\mathbb{T}_{\lambda \lambda }$ only involves
terms that connect real leads, $\mathbb{T}_{\phi \phi }$ only involves
transmissions between decoherent channels and, finally, the blocks $\mathbb{T%
}_{\lambda \phi }$ and $\mathbb{T}_{\phi \lambda }$ connect leads with
decoherent processes. Thus, both $\lambda $ and $\phi $ subscripts may be
vectors themselves indicating processes (current leads $\ell $ or dephasing
processes $\phi $) and states in the system ($n=1,...N$). For instance, for
a system with a single resonant state identified as 1 coupled to two
terminals and a single decoherent process, $\lambda =(L1,R1)$ and $\phi
=\phi 1$. The fact that on-site chemical potentials at decoherent channels
ensure that no net current flows through them, allows us to evaluate $%
\overrightarrow{\delta \mu }_{\phi }$, from Eq. \ref{eq:CurrentsMatrix}: 
\begin{equation}
\overrightarrow{\delta \mu }_{\phi }=\left[ -\mathbb{T}_{\phi \phi }\right]
^{-1}\mathbb{T}_{\phi \lambda }\overrightarrow{\delta \mu }_{\lambda }.
\end{equation}%
Here, $\overrightarrow{\delta \mu }_{\phi }$ provides the chemical potential
profile at the sites undergoing decoherence. Notice that, if used in a local
space representation, these chemical potentials do not distinguish left from
right going electrons. Thus they induce momentum relaxing decoherence. \cite%
{GLBE1,Gasparian96,Datta2007}

The decimative procedure involves a simple algebraic relation between the
real channels of the system and the chemical potentials associated with
currents drains or sources. From Eq. \ref{eq:CurrentsMatrix}, it is
straightforward to isolate $\overrightarrow{I}_{\lambda }$, arriving to the
expression:%
\begin{equation}
\overrightarrow{I}_{\lambda }=\frac{e}{h}\widetilde{\mathbb{T}}_{\lambda
\lambda }\overrightarrow{\delta \mu }_{\lambda },  \label{eq:Idec_matrix}
\end{equation}%
and therefore, the adimensional effective conductances are the non-diagonal
elements of the matrix 
\begin{equation}
\tilde{\mathbb{T}}_{\lambda \lambda }=\mathbb{T}_{\lambda \lambda }+\mathbb{T%
}_{\lambda \phi }[-\mathbb{T}_{\phi \phi }]^{-1}\mathbb{T}_{\phi \lambda },
\label{eq:Teff_matrix}
\end{equation}%
where the first term represents the coherent transmissions while the second
involves all the possible transmissions undergoing at least one decoherent
process. This last term, involves the inversion of a typically big $N\times
N $ matrix. Notice that the matrix in square brackets would correspond to $%
\mathbb{W}$ in the original D'Amato and Pastawski's paper, see Eq. \ref%
{eq:Teff_DP}. \cite{Damato-Pastawski} However, the matrix inversion can be
performed resorting to a recursive decimation of the ${N}$ dephasing
channels, taken one by one. Starting from the first one, at each stage of
decimation, all the remaining probes and dephasing channels become
renormalized according to the following recursive scheme for the matrix
elements of $\tilde{\mathbb{T}}$: 
\begin{eqnarray}
\tilde{T}_{ij}^{\left[ 0\right] } &=&T_{ij}^{{}}  \label{Eq:Trans_order_0} \\
\tilde{T}_{ij}^{\left[ k\right] } &=&\tilde{T}_{ij}^{\left[ k-1\right] }+%
\tilde{T}_{i,k}^{\left[ k-1\right] }\frac{-1}{\tilde{T}_{k,k}^{\left[ k-1%
\right] }}\tilde{T}_{k,j}^{\left[ k-1\right] }.  \label{Eq:Trans_recursive}
\end{eqnarray}%
Here, $k$ runs over the dephasing channel index $\phi {1}...\phi {N}$ and $%
\tilde{T}_{ij}^{\left[ k\right] }$ stands for the matrix element ${i,j}$ \
(each of them take the values $\{\ell {1}...\ell {M,}\phi {1,}...,\phi {N}%
\}) $ of matrix $\mathbb{T},$ after the decimation of $k$ incoherent
channels. This recursion algorithm could become particularly useful when
only the effective transmission among a few external channels is needed.

Once that all of them were decimated, we have an effective transmission
matrix $\tilde{\mathbb{T}}\equiv \tilde{\mathbb{T}}^{(N)}$ given by: 
\begin{equation}
\widetilde{\mathbb{T}}=\left[ 
\begin{array}{cccc}
\tilde{R}_{\ell 1}-1 & \tilde{T}_{\ell 1,L2} & \cdots & \tilde{T}_{\ell
1,\ell M} \\ 
\vdots & \vdots & \ddots & \vdots \\ 
\tilde{T}_{\ell M,\ell 1} & \tilde{T}_{\ell M,\ell 2} & \cdots & \tilde{R}%
_{\ell M}-1%
\end{array}%
\right]  \label{eq:Weff-matrix}
\end{equation}%
which accounts for the overall (coherent plus incoherent) transmission
through the system between different current channels. This effective
transmission matrix relates real currents on each site of the sample with
the voltages associated with each electron reservoir. It should by noticed
that sums over rows or columns, both on the original $\mathbb{T}$ and on $%
\tilde{\mathbb{T}},$ must be zero, in accordance to the Kirchhoff law.

At this point there is a particular situation that should be discussed: a
unique voltage difference between two channel sets. This results in a single
chemical potential difference. For example, assuming that all the channels
associated with a current source in the \textquotedblleft
left\textquotedblright\ source $L$ have the same chemical potential, $\delta
\mu _{L}$ and all those in the current sink $R,$ have $\delta \mu _{R}$. We
can rewrite the net current as: 
\begin{eqnarray}
\mathtt{I} &=&\sum\limits_{i}I_{i}=\frac{e}{h}\sum\limits_{j}^{M_{R}}\sum%
\limits_{i}^{M_{L}}\tilde{T}_{Rj,Li}(\delta \mu _{L}-\delta \mu _{R})
\label{eq:Curr_2Volt} \\
&=&\mathrm{Tr}\left[ 4\boldsymbol{\Gamma }_{R}\mathbb{G}_{N1}^{R}\mathbf{%
\Gamma }_{L}\mathbb{G}_{1N}^{A}\right] (\delta \mu _{L}-\delta \mu _{R})
\label{Eq. Tr(ImGImG)} \\
&=&\mathtt{GV}  \label{eq:GV}
\end{eqnarray}%
where $\mathtt{G}$ is the effective conductance, $\mathtt{V}=(\delta \mu
_{L}-\delta \mu _{R})/e$ is the applied voltage. Notice that $\Gamma _{L}$
and $\Gamma _{R}$ are square matrices with dimensions $M_{L}\times M_{L}$
and $M_{R}\times M_{R}$ associated with the $M=M_{L}+M_{R}$ quantum channels
at the leads $L$ and $R$. Since the final expression is the trace of a
matrix product, the result does not depend on the chosen basis.

For the most general case of several chemical potentials, Eqs. \ref%
{eq:Curr_2Volt}-\ref{eq:GV} can not be used and one should rely on Eqs. \ref%
{eq:Idec_matrix} and \ref{eq:Teff_matrix} that are the general solution to
the multi-terminal DP model. These are the main results of this work
together with the algorithms for the Green's functions, Eqs. \ref%
{eq:thoulessL} and \ref{eq:thoulessR}, and for the effective transmittances,
Eqs. \ref{Eq:Trans_order_0} and \ref{Eq:Trans_recursive}. All of them will
be tested in physically relevant situations in the next two sections.

\section{\label{sec:example1} Application: Decoherence in a Model for a SASER%
}

The explicit description of vibrational degrees of freedom in a transport
problem requires a multichannel formulation even in a two probe
configuration. This is because one must resort to a Fock-space
representation of the Hamiltonian describing electrons and phonons. This
situations occur in vibrational spectroscopy\cite{Stipe98,Park}, 
polaronic models,\cite{BoncaTrugman,BoncaTrugman97} photon-assisted tunneling 
\cite{Stafford96,Jauho98} as well as in time-dependent classical electromagnetic fields
in Floquet representations.\cite{LuisAPL11} 

We will analyze a simple model that represents this family of problems:
independent electrons tunneling through a resonance where they are strongly
coupled to a quantized vibrational mode. In particular, we describe the
optical phonon-assisted tunneling in a double barrier device. It manifests
as a satellite peak in the I-V curve. This mechanism led to one \cite%
{Foa2001} of the various proposals for a phonon laser (SASER).\cite{Kent2010}
In such proposal, a substantial part of the electrons contributing to the
current emit an optical phonon. This constitute the basis for a coherent
ultrasound source. \cite{ChemPhys2002,Camps2001}. The efficiency of the
device depends on the contrast between the satellite peak and the valley,
which in turn is determined by specific quantum interferences among the
participating channels. Thus, we will explore if these interferences survive
the decoherence induced by the acoustic phonons. 
\begin{figure}[tbph]
\begin{center}
\includegraphics[width=2.8in]{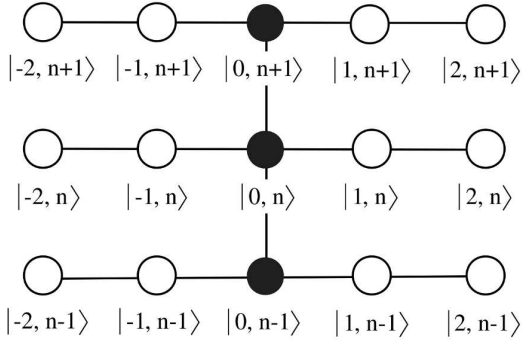}
\end{center}
\caption{Fock-space representation of states $|j,n\rangle $. The middle row
represents local electronic states $j$ with $n$ phonons. Lower and upper
rows describe the same electronic tight-binding chain but with different
numbers of phonons. Vertical lines are local electron-phonon couplings
restricted to site 0th. }
\label{Gr:FockSpace}
\end{figure}

\textit{Model.} Consider a \textquotedblleft local\textquotedblright\
electronic resonant state labeled as $0$. There, the electron is coupled to
a single vibrational mode, with frecuency $\omega _{0}$, whose occupation is
associated with the bosonic number operator $\hat{b}^{\dagger }\hat{b}$.
This is represented by the electron-phonon Hamiltonian,%
\begin{equation}
\hat{H}_{S}=E_{0}\hat{c}_{0}^{+}\hat{c}_{0}^{{}}+\left( \hbar \omega _{0}+%
\tfrac{1}{2}\right) \hat{b}^{+}\hat{b}^{{}}+V_{g}(\hat{b}^{+}+\hat{b}^{{}})%
\hat{c}_{0}^{+}\hat{c}_{0}^{{}}.
\end{equation}%
The eigenstates of this Hamiltonian are the polaron states,\cite%
{BrazJP,Wingreen1988} whose eigenenergies are 
\begin{equation}
E_{0,n}=E_{0}+\hbar \omega _{0}\left( n+\frac{1}{2}\right) -\frac{|V_{g}|^{2}%
}{\hbar \omega _{0}}.
\end{equation}%
The electrons can jump \textit{in} and \textit{out} the resonant state to
the left and right leads. They can also suffer decoherent processes with a
rate $2\Gamma _{\phi }/\hbar $ in a FGR approximation. The effective
Hamiltonian results: 
\begin{equation}
\hat{H}_{\mathrm{eff}}=\hat{H}_{S}+\hat{\Sigma}_{L}+\hat{\Sigma}_{R}+\hat{%
\Sigma}_{\phi },
\end{equation}%
where $\hat{\Sigma}_{L}$ and $\hat{\Sigma}_{R}$ describe the escape to the
current leads and $\hat{\Sigma}_{\phi }$ the escape associated with
decoherence. They are, 
\begin{equation}
\hat{\Sigma}_{L}+\hat{\Sigma}_{R}+\hat{\Sigma}_{\phi }=\left[ \Sigma
_{L}(\varepsilon )+\Sigma _{R}(\varepsilon )-\mathrm{i}\Gamma _{\phi }\right]
\hat{c}_{0}^{+}\hat{c}_{0}^{{}}.
\end{equation}%
Notice that, these self-energies must account for the high voltage
difference required by SASER operation as an offset in the band centers of
the left and right leads $E_{L}-E_{R}=e\mathtt{V}$. We have omitted a real
part of the decoherent process which is not relevant in the present case. As
discussed before \cite{ChemPhys2002}, the optical phonon absorption and
emission can be viewed as a \textquotedblleft vertical\textquotedblright\
processes in a two-dimensional network. Thus, transport in the Fock space is
computationally equivalent to a tight-binding model with an expanded
dimensionality, as shown in Fig. \ref{Gr:FockSpace}.\cite%
{ChemPhys2002,BrazJP,BoncaTrugman} 
\begin{figure*}[tbph]
\begin{center}
\includegraphics[width=5.5in]{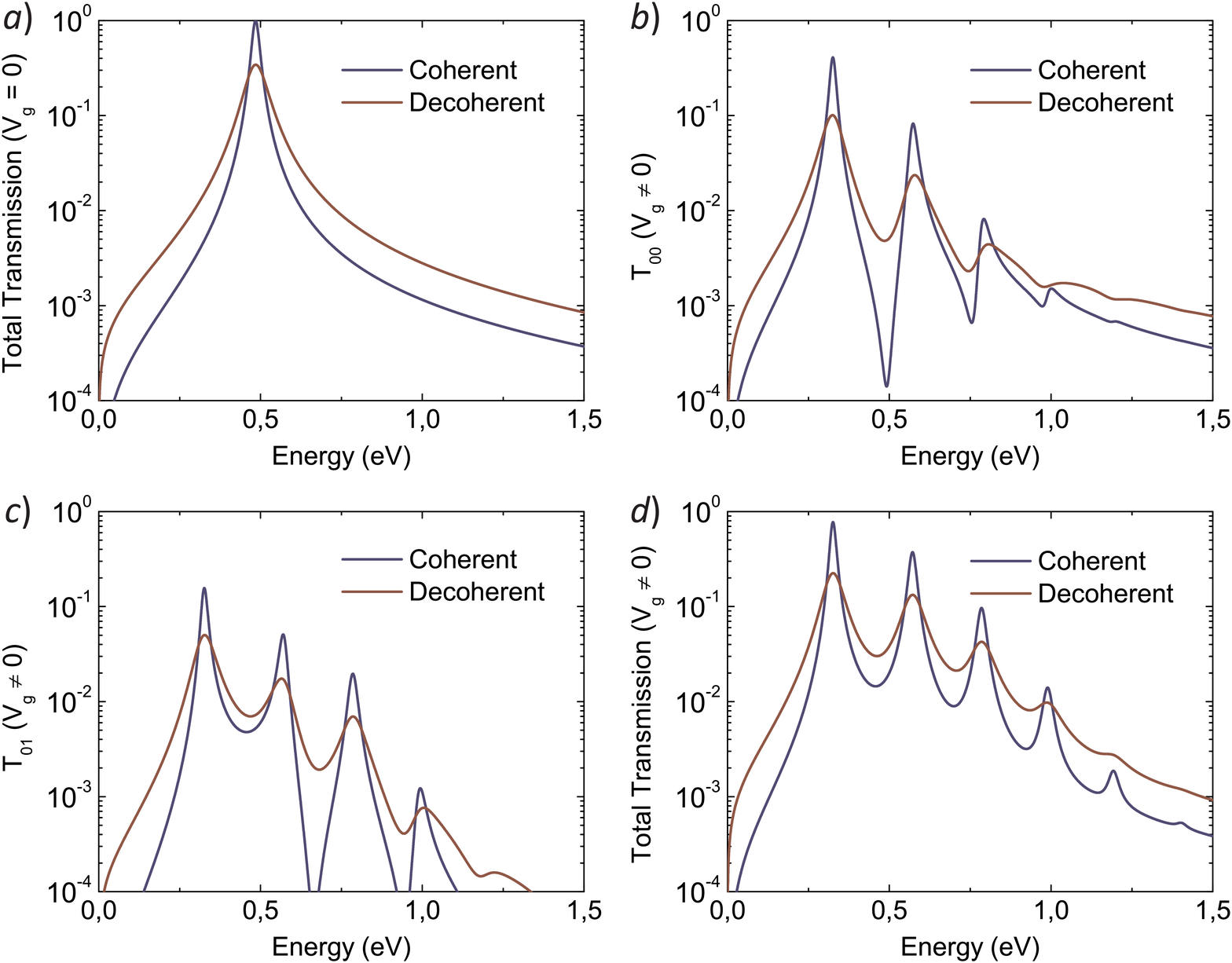}
\end{center}
\caption{Multichannel decoherent transmission for the polaron model, with $%
\hbar \protect\omega _{0}=0.2$ eV, $E_{0}=-1.5$ eV. (a) Local electronic
state without coupling with the phonons ($V_{g}=0$); (b) Transmission
probability for an electron leaving the sample without a change in the
phonon state ($V_{g}=0.1$); (c) Transmission probability for an electron
that leaves the sample emitting one phonon ($V_{g}=0.1$); (d) Total
decoherent transmission probability}
\label{Gr:all}
\end{figure*}

When an electron comes from the left side, it arrives at the resonant site
where it couples to the $n_{0}$ phonons present in the well. It can either
keep its original kinetic energy $\varepsilon -\left( n+\tfrac{1}{2}\right)
\hbar \omega _{0}$ or change it by emitting or absorbing $\Delta n$ phonons.
Thus, the transmission probabilities of each contribution are given by: 
\begin{equation}
T_{R(n_{0}+\Delta n),Ln_{0}}^{{}}=2\Gamma _{R(n_{0}+\Delta
n)}G_{n_{0}+\Delta n,n_{0}}^{R}2\Gamma _{Ln_{0}}G_{n_{0},n_{0}+\Delta n}^{A}.
\label{eq:T_eph}
\end{equation}%
Notice that the subscripts represent channels in the Fock space. As a
consequence of the trivial energy shift, associated with the presence of
phonons, 
\begin{equation}
\Gamma _{\alpha n}(\varepsilon )=\Gamma \left( \varepsilon -E_{\alpha
}-\left( n+\tfrac{1}{2}\right) \hbar \omega _{0}\right) ,
\end{equation}%
for $\alpha =L,R,$ as defined in Eqs. \ref{Dyson}-\ref{self_energy}. Voltages are accounted
by $E_{\alpha }$. Each of this processes contributes to the total coherent
transmission which is given by, 
\begin{equation}
T_{RL}(\varepsilon )=\sum\limits_{\Delta n=-n_{0}}^{\infty }T_{n_{0}+\Delta
n,n_{0}}(\varepsilon ).
\end{equation}%
In an actual device, the current would be obtained integrating $\varepsilon $
with the appropriate Fermi functions. Here, we might recall that Ref. \cite%
{Emberly2000} suggested that in the Fock space, \textquotedblleft
vertical\textquotedblright\ hoppings could be blocked by the presence of
other electrons arriving with different initial energies. However, when the
kinetic energy of the incoming electrons satisfies $E_{F}\leq \hbar \omega
_{0}\leq e\mathtt{V}$, the applied voltage always enables phonon emission 
\cite{ChemPhys2002,BrazJP,BoncaTrugman} ruling out the eventual problem of
overflow \cite{Cattena2012} ensuring the physical significance of our model.

The decoherence is induced by the finite lifetime for the polaron states
through an imaginary correction in the self-energies of Eq. \ref%
{sigma-decoher}. The available \textquotedblleft direct\textquotedblright\
channels are associated with the transmission probabilities of Eq. \ref%
{eq:T_eph}. Because of the wide band approximation for the dephasing
channels, the energy uncertainty is independent of $\varepsilon $:%
\begin{equation}
\Gamma _{\phi n}(\varepsilon )\equiv \Gamma _{\phi }.
\end{equation}%
Optical phonon emission or absorption processes give rise to decoherent
processes, even when $\Gamma _{\phi }=0$. This leaves us with several
possible dephasing channels, whose transmittances are%
\begin{equation}
T_{\beta (n_{0}+\Delta n),\alpha n_{0}}^{{}}=2\Gamma _{\beta \left(
n_{0}+\Delta n\right) }(\varepsilon )|G_{n_{0}+\Delta n,n_{0}}^{R}(\epsilon
)|^{2}2\Gamma _{\alpha n_{0}}(\varepsilon ).  \label{eq:T_eph_dec}
\end{equation}%
Here $\alpha $,$\beta $ are either$~R,L$ or $\phi $. From these
transmissions, and using Eqs. \ref{eq:Teff_matrix} and \ref{eq:Weff-matrix},
we obtain the effective transmissions through the available real channels.
Instead of using the SASER operation regime ($n_{0}\gg 1$), for pedagogical
reasons we will assume that injected electrons find $n_{0}=0$ phonons, a
situation that describes a vibrational spectroscopy experiments. Then the
total transmission is simply, 
\begin{equation}
\tilde{T}_{LR}(\varepsilon )=\sum\limits_{n=0}^{N}\tilde{T}%
_{LR}^{(n)}(\varepsilon ),
\end{equation}%
where each $\tilde{T}_{LR}^{(n)}$ includes the decimation of the incoherent
channels as in Eq. \ref{eq:Decim-T}. In what follows we will analyze $\tilde{%
T}_{LR}(\varepsilon )$ which is also the relevant quantity to study the
non-linear response (see Eq. 134 in Ref. \cite{Pastawski-Medina}).

The total transmission as function of energy is shown in Fig. \ref{Gr:all}.
The Hamiltonian parameters are roughly representative of a double-well
resonant tunneling devices where electron-phonon interactions manifest as a
satellite peak in the conductance. \cite{Foa2001} There $E_{0}=-1.5$ eV, $%
V_{R}=V_{L}=-0.1$ eV, $\hbar \omega _{0}=0.2$ eV and $V_{g}\simeq -0.1$ eV.
We discriminate among different vertical processes contributing to the total
transmittance. When the coupling between the local electronic state and the
phonon mode is neglected, $V_{g}=0$, the problem becomes one dimensional
with a unique resonance, as shown in Fig. \ref{Gr:all}-a. The effect of the
environment, accounted with the DP model, is a broadening of the original
resonance. When the local electronic state is strongly coupled with the
phonon field, $\left\vert V_{g}\right\vert \gg 0$, there are extra available
paths for the conduction electrons in the Fock space. Different electron
pathways in the coherent picture can interfere destructively, e.g. those
that traverse the resonance straight away and those that previously emit and
absorb a virtual phonon. These give rise to anti-resonances in Figs. \ref%
{Gr:all}-b and \ref{Gr:all}-c. Since they are a coherent phenomena, they may
be destroyed when decoherent events are present. This is made evident in
Fig. \ref{Gr:all}-d where the total electron transmission probability in a
multi-phonon process is compared with the same configuration with added
decoherence, according to the multi-terminal DP model.

\begin{figure}[tbph]
\begin{center}
\includegraphics[width=3.4in]{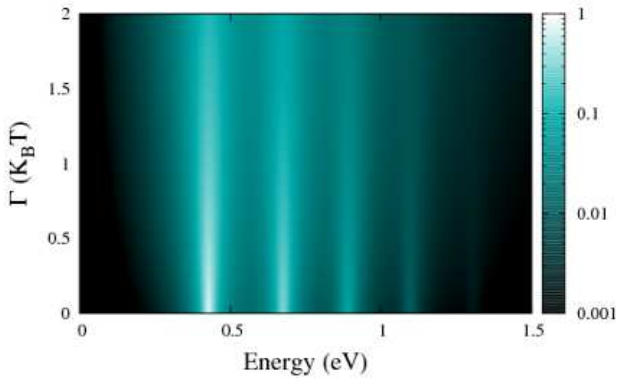}
\end{center}
\caption{Multichannel decoherent transmission for the polaron model in a
color map. The transmission probability is shown in a color scale, as a
function of the incident electron Fermi energy and the strength of the
imaginary energy shift $\Gamma _{\protect\phi }$. The behavior of $\tilde{T}$
is shown to be a smooth function of $\Gamma _{\protect\phi }$.}
\label{Gr:colormap}
\end{figure}

The energy uncertainty used is $\Gamma _{\phi }=0.026$ $e$\texttt{V} $\sim
k_{B}T_{R}$, where $k_{B}$ is the Boltzmann constant and $T_{R}$ stands for
room temperature of $300K$. Although one might evaluate $\Gamma _{\phi }$
from the electronic energy uncertainties obtained with the help of ab-initio
computations, the behavior of $\tilde{T}$ as a function of $\Gamma _{\phi }$
is smooth, provided that these local uncertainties are small compared with
typical tunneling rates from the local resonances, $\Gamma _{L(R)}\gg \Gamma
_{\phi }$. Therefore, small variations of the precise value of $\Gamma
_{\phi }$ do not change the general behavior of $\tilde{T}$. This is
illustrated in Fig. \ref{Gr:colormap} where a color map shows how $\Gamma
_{\phi }$ affects the total transmission probability in the range [$0$ eV,$%
0.025$ eV]. 
We confirmed the general trend that decoherence broadens and lowers the
resonance peaks and raise the tails. But more importantly, valleys are
shaped by multi-phonon coherent processes that produce anti-resonances.
These resulted very sensitive to decoherence. Thus, these effects should be
considered in assessing the efficiency of a SASER.

\section{\label{sec:example2} Application: Quantum to Classical transition
in a Model for Giant Magnetoresistance.}

Spintronics often requires to distinguish how each spin projection
contribute to the current and to identify the spin dependent voltage
profiles, i.e. the chemical potentials $\delta \mu $'s. These are absent
from the original solution of the DP model that just provides the total
current, $I_{LR}=(e/h)T_{eff}\delta \mu $ (see section \ref{sec:DP} ). This
limitation was overcomed by the previous sections, where a specific current $%
I_{j}$, at spin-channel $j$, can be readily calculated from eq. \ref%
{eq:CurrentsMatrix}, as $I_{j}=e/h\sum_{i}\left( \mathbb{T}\right)
_{ji}\delta \mu _{i}$.

Spin-dependent electron transport in ferromagnetic metals presents high
rates of scattering events that could make a fully coherent treatment
somewhat unrealistic. The standard approach is to use the semiclassical
Boltzmann equation \cite{Valet-Fert}. However, in these models quantum
mechanic effects are completely neglected from the very beginning. These
effects can become important and interesting to study. For instance, ref. 
\cite{FernAlcPast13} shows that spin-dependent transmittances in nanowires
with a modulated magnetic field may present Rabi oscillations. In these
situations, a Hamiltonian model capable of reaching a semiclassical limit,
such as the DP, can be very useful.

In this section, we use the multi-terminal DP model to treat one of the
paradigmatic phenomena of the spintronics, the Giant Magnetoresistance
(GMR). We will show that one can go from a purely quantum regime, described
by a Hamiltonian, to the (semi)classical limit of GMR, just by varying a
single parameter: the `decoherent' scattering rate.

Giant Magnetoresistance may occur in systems composed of two layers of a
ferromagnetic metal where their relative magnetization can be switched. In
these materials, the rate of scattering depends on the electron spin. Thus,
the electrical resistance depends on the relative orientation between the
spin and the layer's magnetization. If the two layers have their
magnetization aligned, there is a spin orientation with low resistance that
dominates transport. On the other hand, when the magnetizations are
anti-aligned, both spin channels have high resistance. \cite{Fert08} 

\textit{Model}. Let us consider a one-dimensional system composed of two
adjacent `layers' or portions of a single-domain ferromagnetic metal. We
choose the relative magnetization in a anti-aligned configuration (Fig. \ref%
{fig:ChemPot}-$a$). This system is connected to non-magnetic contacts at
each side, labeled by $L$ and $R$. Thus, the current flows perpendicular to
the magnetic interface. As usual, each spin is regarded as an independent
channel at the contacts. Thus, at the leads, each spin projection is
characterized by the chemical potentials $\mu _{L\uparrow }$, $\mu
_{L\downarrow }$, $\mu _{R\uparrow }$, and $\mu _{R\downarrow }$. Since we
are considering non-ferromagnetic contacts, the chemical potentials at the
leads are spin independent.

Inside the system, the electrons undergo scattering processes producing the
spin dependent resistance. Since its fair to neglect Anderson localization,
we can use the equivalence between delta function impurities and local
decoherent scattering processes. As in the Ohmic limit of the DP model \cite%
{Damato-Pastawski,GLBE1} they can be characterized by the parameter $\Gamma
_{\phi }$. This is related with the mean free time, $\tau _{\sigma }$,
through $\Gamma _{\sigma }=\hbar /(2\tau _{\sigma })$. Then, the Ohmic
conductance is proportional to the mean free path $\ell _{\sigma }$ which
results $\ell _{\sigma }=v_{F}\tau _{\sigma }$. Note that the rate $\Gamma
_{\sigma }$ depends on the relative orientation between the spin and the
local magnetization. Thus, spin $\uparrow $ has a scattering rate $\Gamma
_{\phi 1}$, at the first layer, and $\Gamma _{\phi 2}$, at the second one.
The opposite spin has the complementary rates.

\begin{figure}[tbh]
\begin{center}
\includegraphics[width=3.4in]{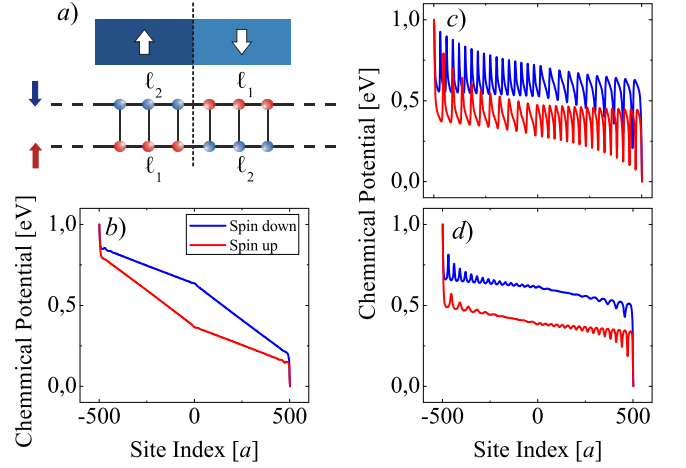}
\end{center}
\caption{-$a)$ On top is a scheme showing the layer's magnetization in the
two resistor model for GMR. Below is a tight binding representation
discriminating the spin projection. The coherence lengths of electrons in
the first layer are $\ell _{1}$ and $\ell _{2}$ for up and down spin
electrons respectively. Note that coherence lengths are inverted in the next
layer. $\ell _{1}/\ell _{2}=1/2$ in all cases. Fig. $b)$ to $d)$ Site
dependent chemical potentials with $\ell _{1}=15~a$ in Fig. $b)$, $\ell
_{1}=1500~a$ in Fig. $c)$, and $\ell _{1}=150~a$ in Fig. $d)$. The system
length is $1000a$ and $V_{\downarrow \uparrow }=0$ ($L_{sd}\rightarrow
\infty $), and the chemical potentials at the leads are $\protect\mu _{L}=e%
\mathrm{V}$ and $\protect\mu _{R}=0$. The Fermi wavelengths at the left side
are $\protect\lambda _{F}=45a$, for up spins, and $\protect\lambda _{F}=30a$%
, for down spins. The opposite holds at the right ferromagnet. The chosen
parameters do not represent a specific experimental set up.}
\label{fig:ChemPot}
\end{figure}

As in previous works,\cite{Gopar,FernAlcPast13} the system's Hamiltonian $%
\hat{H}_{S}$ is described in a tight-binding approach that includes local
spin-reversing interactions: 
\begin{eqnarray}
\hat{H}_{S} &=&{\sum\limits_{i=-N}^{N}}\sum\limits_{\sigma =\uparrow
,\downarrow }[E_{i,\sigma }^{{}}\hat{c}_{i,\sigma }^{\dag }\hat{c}_{i,\sigma
}^{{~}}+V\left[ \hat{c}_{i\sigma }^{\dagger }\hat{c}_{i+1\sigma }^{{~}}+%
\mathrm{c.c.}\right]  \notag \\
&&+{\sum\limits_{i=-N}^{N}}V_{\downarrow \uparrow }\left[ \hat{c}%
_{i\downarrow }^{\dagger }\hat{c}_{i\uparrow }^{{~}}+\mathrm{c.c.}\right] .
\label{HamGMR}
\end{eqnarray}%
The label $i$ indicates sites on a lattice with unit cell $a$, $E_{i,\sigma
}^{{}}$ is the energy at the site $i$ with spin $\sigma $, the operator $%
\hat{c}_{i,\sigma }^{\dag }$ ($\hat{c}_{i,\sigma }^{{~}}$) creates
(annihilates) a particle at the site $i$ with spin $\sigma $. The firsts two
terms of $\hat{H}$ accounts for the site energies and the spin-conserving
hopping, $V$, between adjacent sites. $V$ is chosen as the unit of energy.
In a graphical representation, each spin orientation is represented by a
chain of sites interconnected by $V$. Thus, two chains of sites are needed
to represent the spin-dependent transport along this ferromagnetic system
(Fig. \ref{fig:ChemPot}-$a)$). The last term of $\hat{H}$, models the
scattering processes that may change the spin projection, such as scattering
with magnetic impurities. Thus, $V_{\downarrow \uparrow }$ is the local
spin-reversing or spin-mixing hopping parameter. This is related to a
characteristic length scale identified as the spin diffusion length, $L_{sd}$%
, by 
\begin{equation}
L_{sd}=\frac{\hslash v_{F}}{2\left\vert V_{\downarrow \uparrow }\right\vert }%
,
\end{equation}%
where $v_{F}$ is the Fermi velocity and $L_{sd}$ is the length scale at
which the spin-flipping processes relax the diffusing spin. Thus, within
this length, both spin orientations can be considered as independent. $%
L_{sd} $ is typically much larger than the mean free path. When the electron
gets into the ferromagnetic material, it undergoes an \textit{exchange
interaction} that can be regarded as a Zeeman interaction. Thus, the site
energy is $E_{i,\uparrow (\downarrow )}=E_{0}\pm \Delta E_{Z}$, where $i$ is
a site of the first layer.

As in Eq. \ref{Hamil-efectivo}, the effective Hamiltonian incorporates the
leads and the scattering processes through the appropriate self-energies.
Now, $\hat{\Sigma}_{L(R)}=\hat{\Sigma}_{L(R)\uparrow }+\hat{\Sigma}%
_{L(R)\downarrow }$ is the self-energy operator describing the escape to the
left (right) lead, given by Eq. \ref{Dyson}, where all hoppings are equal to 
$V$. Decoherent channels accounting for resistive scattering are associated
to each site and included into $\hat{H}$ through the $\hat{\Sigma}_{\phi }$
operator. Thus, $\hat{\Sigma}_{\phi }$ is diagonal in a matrix
representation. In the wide band limit, their elements are purely imaginary,
i.e. $\left( \hat{\Sigma}_{\phi }\right) _{ii}=-\mathrm{i}\Gamma _{\phi i}$.

\textit{Classical regime of GMR: two resistors model (TRM)}. Here, the
system length is much shorter than $L_{sd}$, i.e. $V_{\downarrow \uparrow
}\approx 0$ in Eq. \ref{HamGMR}. Here, when electrons enters into a
ferromagnetic layer they undergo an electrical resistance $\delta R=I_{LR}%
\mathrm{V}$ (Ohm's law) that manifest in a linear drop in the
chemical-potential $\delta \mu $. Therefore, in the anti-aligned
configuration, there are two linear potential drops of $\delta \mu $ with
slopes proportional to the spin-dependent resistance of each layer. Then, it
is expected a splitting of the chemicals potentials that form a diamond like
figure. This is precisely what we obtain using the multi-terminal DP method
with mean free paths shorter that the system size. Fig. \ref{fig:ChemPot}-$b 
$ to \ref{fig:ChemPot}-$d$ show this, through the site-dependent chemical
potential. In contrast, for the quantum limit of long mean free paths,
quantum interferences are evident. However, they are smoothed out by
increasing the scattering rate until they reach the expected classical
diamond like figure (Fig. \ref{fig:ChemPot}-$b$). 
\begin{figure}[tbh]
\begin{center}
\includegraphics[width=2.5in]{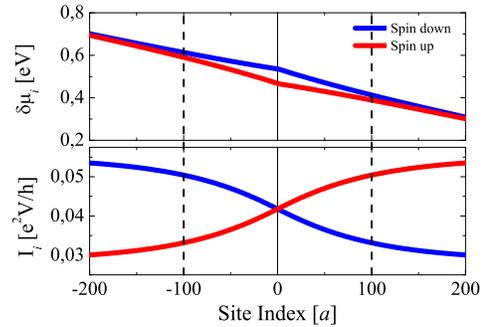}
\end{center}
\caption{Upper figure, site-dependent chemical potential $\protect\delta 
\protect\mu _{i}$ profile for the semiclassical model of GMR with finite
spin diffusion length, $L_{sd}=100~a$. Lower figure, shows the local
currents $I_{i}$ for up and down spin electrons. System size is $1000a$, $%
\ell _{1}/\ell _{2}=1/2$, and $\ell _{1}=15a$.}
\label{fig:diamond}
\end{figure}

\textit{Semiclassical regime of GMR: Valet and Fert theory.} Considering
finite values for the spin diffusion length, $L_{sd}$, Valet and Fert \cite%
{Valet-Fert} showed that the difference of the spin-dependent local chemical
potentials decays exponentially with the distance to the magnetic interface
with a length scale given by $L_{sd}$. They also showed that the
spin-dependent current is inverted in this length scale. In Fig \ref%
{fig:diamond} we show that the multi-terminal DP model is also capable to
reproduce these behaviors provided that we turn on the spin flip term in Eq. %
\ref{HamGMR}. In the upper figure we show the spin and site dependent
chemical potentials. One can see that in regions far from the interface,
distances larger than $L_{sd}$, the chemical potentials are nearly the same.
In regions close to the interface, the chemical potential drop forms a
diamond-like figure that show the expected spin-dependent exponential
contributions summed up to the trivial mean linear drop. In the lower figure
we can observe how the inversion of the currents is produced in the length
scale $L_{sd}$. For longer distances, the currents reach a stationary value.

All these behaviors are in agreement with Ref. \cite{Valet-Fert}. This
situation reinforces the descriptive conceptual value of the DP model and
the versatility of the numerical algorithms developed in this paper.

\section{\label{sec:conclusions}Conclusion}

In this work, we first reviewed the original two-terminal DP model, which
accounts for decoherent effects in quantum transport. Then, we presented an
extension of this model which is capable to deal with multi-terminal setups.
Also, we introduced recursive algorithms that allows us to take advantage of
the problem symmetries, in particular in the case of general banded
Hamiltonians. The incorporation of a unified notation gives more
transparency to its potentialities. Using the specific Hamiltonian models
for phonon laser and giant magnetoresistance, we exemplified how to treat
multi-channel problems in presence of decoherence.

We made special emphasis on the role of decimation procedures in the context
of banded effective Hamiltonians, since they can be used as the basis for
efficient computational schemes. In particular, one of the keys is given by
Eqs. \ref{eq:thoulessR}- \ref{eq:thoulessL}. Note that, in the very common
situation of block tridiagonal (i.e. banded) matrix Hamiltonians, these
recursive equations provide an efficient decimation procedure that allows
one to obtain all the $N\times (N-1)$ non-diagonal blocks of the whole
Green's function matrix, $\mathbb{G}$, in terms of the $N$ diagonal blocks.
In turn, these last can be calculated as matrix continued fractions. \cite%
{MCF-Pastawski} The idea here is to take advantage of particular system's
symmetries using these expressions to build an efficient computation
approach for the problem under study.

Profiting from a parallelism between the computation of $\mathbb{G}$ and the
decoherent transmitance $\tilde{\mathbb{T}}$ already hinted by the DP
solution \cite{GLBE1}, we also derived a compact matrix equation for $\tilde{%
\mathbb{T}}$ in a generalized multi-terminal scheme. This recursive
algorithm relies on decimation procedures.

As a first application, we added decoherent processes to the usual model for
phonon-assisted tunneling in the configuration used for a phonon laser or
SASER. As is well known, \cite{Foa2001} in the I-V curve of a SASER
configuration, the contrast between the valley (out of resonance) and the
satellite peak (corresponding to phonon emission) is enhanced by the effect
of antiresonances. These last result from the interference between different
paths in the Fock's space. \cite{ChemPhys2002} Besides of the expected
smoothing out of the resonances because of the decoherence, we found that it
leads to the degradation of the contrast mainly from the suppression of the
antiresonances. This could set up new bounds for the efficiency of SASER
operation. \cite{Camps2001}

We also solved a simple multi-terminal DP model representative of the giant
magnetoresistance (GMR) phenomenon. There, each spin orientation is a
different conduction channel. Thus, the spin-dependent transport is
intrinsically multi-terminal. We essentially showed that the main
characteristics of the GMR can be well reproduced with this simple model.
While preserving a Hamiltonian description, it is able to reach the expected
classical and semiclassical regimes by means of a single parameter, the
local decoherent rate $\Gamma _{\phi i}$. What is more important, as in Fig. %
\ref{fig:ChemPot}-$c$ and $d$, it opens the possibility to profit from
situations where quantum interference becomes relevant. \cite%
{FernAlcPast13,Richter12}

With increasing system's size, molecular electronics suffers a paradigm
shift on its dominant transport mechanism, from \textquotedblleft coherent
tunneling\textquotedblright\ to \textquotedblleft incoherent
hopping\textquotedblright. Within this context, the present work should
result specially helpful in providing a computational bridge between these
limiting situations, while maintaining a general, transparent, and efficient
approach to quantum transport.

\section{Acknowledgments}

We acknowledge L. E. F. Foa Torres for his comments and stimulating
discussions at an early stage of this work. We received financial support
from ANPCyT, CONICET, MiNCyT-Cor, and SeCyT-UNC.

\bibliography{./MDP-43.bib}

\begin{thebibliography}{10}
\bibitem{SquareRoot}%
  \BibitemOpen
  \bibinfo {note} {The infinitesimal $\eta$ with the chosen sign ensures the
  correct branches of the solution. Note that this expression is easily
  obtained from the definition of the square root of a complex number: \newline
  $\sqrt{x+\mathrm{i}y}=\sqrt{\frac{r+x}{2}} +
  \mathrm{i}\text{sgn}(y)\sqrt{\frac{r-x}{2}}$ where $r = |x+\mathrm{i}y|$.
  \newline Also note that the sign of this infinitesimal is consistent with Eq.
  13.}%
  \bibAnnoteFile{Stop}{SquareRoot}%
\end{thebibliography}

\end{document}